\documentclass[3p,sort&compress]{elsarticle}
\usepackage{amsmath}
\usepackage{bm}
\usepackage{txfonts}
\usepackage[english]{babel}
\usepackage{subfigure}

\journal{Annals of Physics}

\begin{document}

\begin{frontmatter}

\title{Scalar quantum kinetic theory for spin-$1/2$ particles: mean field theory}

\author{Jens Zamanian\corref{cor}} 
\ead{jens.zamanian@physics.umu.se}
\author{Mattias Marklund\corref{cor2}} 
\ead{mattias.marklund@physics.umu.se}

\author{Gert Brodin}
\ead{gert.brodin@physics.umu.se}

\address{Department of Physics, Ume{\aa} University, SE--901 87 Ume{\aa}, Sweden}

\cortext[cor]{Corresponding author}
\cortext[cor2]{Principal corresponding author}

\begin{abstract}
Starting from the Pauli Hamiltonian operator, we derive a scalar quantum kinetic equations for spin-$1/2$ systems. Here the regular Wigner two-state matrix is replaced by a scalar distribution function in extended phase space. Apart from being a formulation of principal interest, such scalar quantum kinetic equation makes the comparison to classical kinetic theory straightforward, and lends itself naturally to currently available numerical Vlasov and Boltzmann schemes. Moreover, while the quasi-distribution is a Wigner function in regular phase space, it is given by a Q-function in spin space. As such, nonlinear and dynamical quantum plasma problems are readily handled. Moreover, the issue of gauge invariance is treated. Applications (e.g.~ultra-dense laser compressed targets and their diagnostics), possible extensions, and future improvements of the presented quantum statistical model are discussed. \\[2mm]
\end{abstract}

\end{frontmatter}


\section{Introduction}
Quantum kinetic theory has a long history. In many respects, it all started with the seminal paper by Wigner in 1932 \cite{wigner}, see also Refs. \cite{weyl,groenewold}, and the later developments of Moyal \cite{moyal}. While the approach of Wigner has the advantage of being of interest for the interpretation of quantum mechanics \cite{zachos}, and also for the development of quantum optics (for an overview, see e.g. \cite{Leonhardt1997}), detailed calculations of material properties in  condensed matter systems have relied to a large extent on either semiclassical techniques \cite{haug-jauho}, in which the collisional operator in Boltzmann's equations involves quantum transition probabilities, or Green's function techniques \cite{baym-kadanoff,kadanoff-baym}, as well as diagrammatic techniques \cite{rammer}. The theory of Baym and Kadanoff, as well as the works of Keldysh \cite{keldysh1,keldysh2}, has been successful in dealing with certain quantum transport phenomena. The theory contains memory effects (nonlocal terms, both in space and time), has a straightforward interpretation in terms of the different Green's functions, and the theory works well even on time-scales shorter than the typical relaxation time of the system in question. However, the gap between classical plasma physics and quantum transport theory does not seem to have been bridged, probably due to reasons of formalism as well as a difference in application of the respective models. Moreover, while the Kadanoff-Baym equation gives a very good description of certain systems, it is perhaps not well-suited to some of the future applications of quantum kinetic theories, such as high intensity laser-plasma interactions \cite{eliezer}, high energy density physics \cite{drake}, and nonlinear collective quantum problems \cite{anderson-etal:2002,marklund:2005,shukla-etal:2006,nshukla:2009,shukla-eliasson2009}. 

In particular, the field of quantum plasmas has recently attracted, a perhaps unexpected, interest in the field of laser plasmas \cite{glenzer,kritcher,lee-etal}, where high density ionized plasmas can be created in the laboratory. Moreover, the event of nano-devices and technology on sub-micron scales, such as quantum dots \cite{alivisatos,haas,manfredi-hervieux} and plasmonic components \cite{maier,epl}, has sparked the interest of many researchers of analyzing the dynamic and nonlinear properties of such systems. A recent result is that quantum effects in plasmas can be important in parameter regimes that for a long time have been considered purely classical \cite{Brodin-Marklund-Manfredi}.

The above discussion is mainly related to the statistical and dispersive behavior of unmagnetized quantum plasmas \cite{melrose}. However, one intrinsic non-classical property of quantum systems is the spin. 
The magnetization that follows from the intrinsic spin, as well as that of orbital angular momentum, is of course the foundation for many important material properties \cite{stancil-prabhakar}. Investigations of such condensed matter systems are often directed towards equilibrium properties, although the nonlinear dynamics of magnetization is sometimes interest and probed using the Landau-Lifshitz-Gilbert equation  \cite{bertotti-etal}. 
There are a variety of different physical systems where the spin can be of importance, such as metal alloys and semiconductors material for memory use  \cite{bertotti-etal}, cold atom gases \cite{pethick}, and high density and high field astrophysical plasmas \cite{drake}, to mention a few.
Collective effects originating in the plasma particle species spin has therefore recently become an active field of research for fully ionized systems (see e.g. \cite{marklund-brodin2007,shukla-eliasson2009} and references therein), in particular in the nonlinear regime, where spin solitons \cite{brodin-marklund_pop:2007} and ferromagnetic behavior in plasmas can be found \cite{brodin-marklund_pre:2007}. Many of the studies presented in the literature have so far been of a theoretical nature, but it is not difficult to envision future applications to e.g. plasmonic devices \cite{maier} or femtosecond physics \cite{grossman}.

For the purpose of connecting classical plasma physics to the evolution of nonequilibrium quantum systems, utilization of quasi-distributions is of great value. First of all, the interpretation of the quasi-distribution function using ensemble averages of observables is in direct analogy with the classical case. It is even possible to directly construct a quasi-distribution, such as the Wigner function from measurements \cite{kurtsiefer} (with the only information loss being the initial phase). Second, the quasi-distribution evolution follows from the quantum Liouville equation for the density operator, and gives a quantum analog of the Vlasov or Boltzmann equation. This may also render a quantum kinetic theory for the quasi-distribution function useful for adaption of classical numerical codes to the quantum regime. There are of course infinitely many ways to construct a quasi-distribution function, giving certain elementary requirements (see next section). However, a few quasi-distribution functions are more prominent in the literature than others. 
The best known quasi-distribution function is probably the Wigner distribution \cite{wigner}, but there are many others frequently used. In short, different definitions correspond to different operator ordering, so depending on the application different definitions are natural. For example, when considering optical coherence normally ordered operators occurs naturally and hence the  Glauber-Sudarshan P-distribution \cite{Glauber1963,Sudarshan1963} is a convenient choice. On the other hand, anti-normal ordered operators i.e. the Q-function or the more general Husimi function \cite{Husimi1940} are useful when dealing with quantum chaotic systems. For reviews of the subject see for example Refs.\ \cite{zachos,hillery,lee}. 

In this paper, we will construct a quasi-distribution function for a particle with spin-1/2 as a combination of a Wigner distribution for the position and momenta and the Q-function for the spin degree of freedom. Moreover, a quantum kinetic equation giving the evolution of this scalar distribution function, in the mean field or Hartree approximation, will be derived and applications to magnetized systems will be presented. A discussion of possible future applications and research directions will also be given.  

The structure of the paper is as follows. In Sec.\ 2 we give a short overview of different quantum quasi-distributions. In Sec.\ \ref{sec:1} we consider the evolution equation for a density matrix for a spin-1/2 particle in an external electromagnetic field. In Section \ref{sec:2} we go on to derive a combined transformation for the phase space and spin variable. This transformation then renders an evolution equation for the system in extended phase space $(\mathbf x, \mathbf p, \hat{\mathbf s})$ which is derived in Section \ref{sec:evo}. The extension to the mean field approximation is reviewed in Section \ref{sec:many} and in the following section we calculate the thermodynamic equilibrium density matrix for a set of $N$ noninteracting particles. In Section \ref{sec:3}  we consider the evolution equation in the long scale length limit and compare our results to previous semi-classical kinetic descriptions in the literature. In Section \ref{sec:lin} we consider the linear solutions to the derived equations. Section 10 is devoted to a discussion of gauge properties and the fully gauge invariant evolution equation is presented. Finally we summarize the main results and discuss future development and applications in Section \ref{sec:sum}.

\section{General requirements of quasi-probability distribution function}

\subsection{Historical note}

Following the success of the classical theory of non-equilibrium statistical mechanics, 
it was natural to seek a similar theory for quantum systems in the late 20s and early 30s. 
However, while the classical Liouville equation generates trajectories in phase space as
in a classical Hamilton-Jacobi theory, we in the quantum realm have to consider the 
Heisenberg uncertainty principle. This will not allow us to describe, as in classical systems,
precise trajectories, but rather "smeared out" paths in what would be the corresponding phase space. 
Indeed, the attempts by de Broglie, Bohm, and others to give a close-to classical interpretation of the Schr\"odinger equation by Hamilton-Jacobi theory shows that, if one is inclined to stick to this interpretational scheme and extend this to statistical interpretations, one has to consider the wave function rather as an ensemble of (nonclassical \cite{styer,ballentine}) trajectories (a similar conclusion can be drawn from path integral \cite{dauger-etal} as well as Ehrenfest techniques \cite{ballentine-etal}), satisfying certain initial and boundary conditions. Thus, the introduction by Wigner of a quasi-distribution function (see below) was a natural step in the direction of relating measurements to classical transport theory. This is perhaps most obvious in the field of quantum optics, where phase space techniques since long has been widely used. Three main definitions of quasi-distributions can be found in this field, namely the Wigner function \cite{wigner}, the Husimi (or, equivalently, the Q-) function \cite{Husimi1940}, and the Glauber-Sudarshan P-distribution \cite{Glauber1963,Sudarshan1963}. Below we will give a short summary of some of the properties of the first two types of quasi-distribution functions (the P-distribution will not be used in the present work).

\subsection{Basic requirements}

Some basic requirements can be imposed on a quantum probability distribution function in phase space, in order for it to have a reasonable interpretation \cite{ballentine,Schleich2001}.  
We denote the quantum (quasi-)distribution function by $f(\textbf{x}, \textbf{p})$ (for the moment, we drop the explicit time-dependence for notational convenience) for a given quantum state $\hat \rho$ of the system. Then the marginal distribution functions $\langle \mathbf{x} | \hat \rho | \mathbf{x} \rangle$ and $\langle \mathbf{p} | \hat \rho | \mathbf{p} \rangle$ should be related to $f(\mathbf{x}, \mathbf{p})$ according to 
\begin{equation}\label{eq:marginal_x}
	f(\mathbf{x}) \equiv \int d^3p\,f(\textbf{x}, \textbf{p}) = \langle\mathbf{x} | \hat \rho | \mathbf{x} \rangle ,
\end{equation}
and
\begin{equation}\label{eq:marginal_p}
	f(\mathbf{p}) \equiv \int d^3x\,f(\textbf{x}, \textbf{p}) = \langle\mathbf{p} | \hat\rho | \mathbf{p} \rangle ,
\end{equation}
respectively. Moreover, we should require that the distribution function is positive definite, i.e.\
\begin{equation}\label{eq:positivity}
	f(\textbf{x}, \textbf{p}) \geq 0 .
\end{equation}
However, it can be shown that the conditions (\ref{eq:marginal_x})--(\ref{eq:positivity}) is not sufficient to uniquely determine a suitable quantum distribution function in phase space. In fact, Cohen \cite{Cohen1986} has shown that there are infinitely many function $f(\mathbf{x}, \mathbf{p})$ satisfying (\ref{eq:marginal_x})--(\ref{eq:positivity}). 

A more complete list of properties that are desirable is found in \cite{lee}, where expect for the three properties above, the additional properties that the distribution function is real, bilinear in the wave function and that the distribution functions for eigenstates of the Hamiltonian form a complete and orthogonal set. In fact, it can be shown that in general one cannot find a distribution function that satisfies all of (\ref{eq:marginal_x})--(\ref{eq:positivity}) simultaneously, if one requires the distribution function to be bilinear in the wave function \cite{wigner71}. 

Though the above conditions are important when it comes to interpreting the distribution functions a perhaps more important condition is that it should be possible to calculate the expectation value of any operator. This condition is important since it means that all physically relevant information is included. To calculate the expectation value one  first map the operator to the corresponding phase space function $\hat O = O(\hat{\mathbf x}, \hat{\mathbf p}) \rightarrow O(\mathbf x, \mathbf p)$, using the Weyl-correspondence, and then calculate the phase space average weighted by the distribution function
\begin{equation}
	\langle \hat O \rangle = \int d^3 x d^3 p f(\mathbf x, \mathbf p) O(\mathbf x, \mathbf p) . 
\end{equation}
The mapping from the operator space to phase space depends on which distribution function is used (see Ref.\ \cite{lee} for a details). Below we will collect the properties of two distribution functions of interest in our context, the Wigner distribution \cite{wigner} and the Husimi function \cite{Husimi1940} (or $Q$-function \cite{Leonhardt1997,Schleich2001}). These are also perhaps the most frequently encountered quantum probability distribution function in the literature (see Ref.\ \cite{Leonhardt1997} and \cite{Schleich2001} for further references and other prominent distributions used in quantum optics, such as the $P$-distribution of Glauber \cite{Glauber1963} and Sudarshan \cite{Sudarshan1963}, and their interrelations). 

\subsection{The Wigner function}

The Wigner function for a quantum state $\hat \rho$ is defined as the Fourier transform of the two-point correlation function (i.e., density matrix). Thus, we accordingly have 
\begin{equation}
	f_W(\textbf{x}, \textbf{p}) = \frac{1}{(2\pi\hbar)^{3}} \int d^3 y\, 
		e^{i\mathbf{p}\cdot\mathbf{y}/\hbar}
		\langle \mathbf{x} + \mathbf{y}/2 |\hat \rho | \mathbf{x} - \mathbf{y}/2 \rangle .
\end{equation}
Through this definition of the Wigner function, we see that it satisfies the marginal distribution requirements (\ref{eq:marginal_x}) and (\ref{eq:marginal_p}). However, it does not satisfy the positivity criteria (\ref{eq:positivity}). The latter property then prevents a probability distribution interpretation. However, the negativity of the Wigner function is limited in the sense that the proper number density in physical space is $n(\mathbf{x}) = \int\,d^3 p\,f_W(\mathbf{x},\mathbf{p})$ which is thus always positive. For a pure state $\hat \rho = | \psi \rangle\langle \psi |$, this definition gives
\begin{equation}
	f_W(\textbf{x}, \textbf{p}) = \frac{1}{(2\pi\hbar)^{3}} \int d^3 y\, 
		e^{i\mathbf{p}\cdot\mathbf{y}/\hbar}
		\psi^*( \mathbf{x} + \mathbf{y}/2) \psi(\mathbf{x} - \mathbf{y}/2) .
\end{equation}
One of the important properties of the Wigner function is that it cannot have too sharp peaks, expressed by
\begin{equation}
 	\int\int d^3 x\, d^3p\,[f_W(\mathbf{x}, \mathbf{p})]^2 \leq \frac{1}{(2\pi\hbar)^{3}}  ,
\end{equation}
a result of the noncommutativity between coordinate and momentum operators. 

The time evolution for the Wigner function in an external (analytic) potential $V(\mathbf{x},t)$ is given by
\begin{equation}\label{eq:wigner-evolution}
  	\frac{\partial f_W}{\partial t} + \frac{\mathbf{p}}{m}\cdot\nabla_x f_W 
		+ \frac{2V}{\hbar}\sin\left( 
			\frac{\hbar}{2}\stackrel{\leftarrow}{\nabla_{x}}\cdot\stackrel{\rightarrow}{\nabla}_{p} 
		\right)f_W = 0 ,
\end{equation}
where the $\sin$-function is defined in terms of its Taylor expansion in the case of analytic potentials, and we have used the indices $x$ and $p$ on the $\nabla$ to denote its operation in phase space. 
To find the phase space function that corresponds to a given operator we must first express the operator in Weyl order \cite{weyl}, i.e. express in symmetric products of $\hat x_i$ and $\hat p_i$, $i=1,2,3$, using the commutation relations and then substitute $\hat x_i \rightarrow x$ and $\hat p_i \rightarrow p$.  For example, calculating the average of the operator $\hat x_i \hat p_j$, we have 
\begin{equation}
	\hat x_i \hat p_j = \frac{1}{2} \left( \hat x_i \hat p_j + \hat p_j \hat x_i \right) + \frac{i\hbar}{2} \delta_{ij}
	\rightarrow x p + \frac{i\hbar}{2} \delta_{ij}, 
\end{equation}
where $\delta_{ij}$ denotes the Kronecker delta function, and hence 
\begin{equation}
	\left< \hat x_i \hat p_j \right> = \int d^3 x d^3 p f_W(\mathbf x, \mathbf p) \left( x_i p_j + \frac{i\hbar}{2} \delta_{ij} \right) .
\end{equation}

\subsection{The Husimi function} 

The Husimi function (see (\ref{eq:husimi_def}) below) is based on minimum uncertainty wave packets, and it  
does not satisfy (1) and (2) but is positive definite (thus satisfying (3)). As will be seen below, this allows probability distribution interpretation of the Husimi function; however, it gives a different greater uncertainty measure than expected through naive application of the Heisenberg uncertainty relation. These properties can be immediately understood from the following definition. For a given Wigner function, the Husimi function can be obtained through a Gaussian smoothing as 
\begin{eqnarray}
	f_H(\mathbf{x}, \mathbf{p}) = \frac{1}{\left(\pi\hbar\right)^{3}} \int\int d^3x'd^3p'\,
	\exp[-(\mathbf{x}' - \mathbf{x})^2/2d^2]
	\exp[-\hbar^2(\mathbf{p}' - \mathbf{p})^2/2d^2]
	f_W(\mathbf{x}', \mathbf{p}') ,
\label{eq:husimi_def}
\end{eqnarray}
where the parameter $d$ sets the scale of the smoothing.

While the Husimi function is positive definite, and produce the correct expectation values of observables, it satisfies an indeterminacy relation of the form
\begin{equation}
	(\Delta x)_H(\Delta p)_H \geq \hbar ,
\end{equation} 
as compared to the relation
\begin{equation}
	\Delta x\Delta p \geq \hbar/2 
\end{equation} 
for a quantum state (the latter being satisfied by the Wigner function). This results is due to the smoothing introduced in the definition of the Husimi function. The Husimi function does not give the probability for the particle to be at a certain phase space position, but rather the probability to find the particle in the \textit{minimum uncertainty state} centered around the phase space point in question \cite{harriman}.  Introducing minimum uncertainty states $\left|\mathbf x_0, \mathbf p_0\right>$ which satisfies $\Delta x^2 \Delta p^2 = \hbar^2/4$ one can write the Husimi function as
\begin{equation}
	f_H ( \mathbf x, \mathbf p) = \left< \mathbf x , \mathbf p \right| \hat \rho \left| \mathbf x , \mathbf p \right> .
	\label{husimi}
\end{equation}
However, as mentioned above, it can still be used to calculate any observable, but the operator ordering rule is more complicated than in the Wigner case and we will not consider this further here. The evolution of the Husimi equation can be found from (\ref{eq:wigner-evolution}) and the definition (\ref{eq:husimi_def}). It is fairly complicated (see \cite{lee}) and it is more convenient to compute the evolution of the Husimi distribution function by evaluating the Wigner function for all times through (\ref{eq:wigner-evolution}). This said, we note that although the evolution equation for the Husimi function is more complicated than the corresponding equation for the Wigner equation, it is sometimes the convenient choice. One such example is when considering chaotic system in which the phase space distribution function becomes very complicated. The Husimi function, being a Gaussian average, may then behave more regularly (see, e.g., Refs.\ \cite{lin-ballentine,izrailev,stockmann}). 

\subsection{Quasi-distribution functions for spin}
Similarly to the case for phase space it is possible to construct quasi-distribution functions for the spin degree of freedom. This has been done already in the 1950's by Stratonovic \cite{stratonovic}. Later on the spin quasi-distribution functions were further developed and were applied to problems related to calculating correlation between spins \cite{scully83,cohen,chandler}. The spin quasi-distribution function has also been discussed in connection with quantum scattering problems \cite{carruthers}. 

As in the case of the regular phase space variables $\mathbf{x}$ and $\mathbf{p}$, there is no unique way to introduce a spin quasi-distribution function. Scully and W\'odkiewicz \cite{scully} give a very good review of the many different choices that can be considered. There are at least three different methods for defining spin distribution functions: delta distributions, distributions based on coherent states (Q and P) and Stratonovic distribution functions. However, the different outcomes of these choices overlap. 

In this paper we will consider only the Q-function for spin which is defined as 
\begin{equation}
	f(\theta, \varphi) = \left< \mathbf s \right| \hat \rho \left| \mathbf s \right> , 
	\label{q-spin}
\end{equation}
where $\left| \mathbf s \right>$ is the state which has spin up in the direction of the unit vector $\mathbf s \mathbf = \mathbf s(\theta, \varphi)$ often called a spin-coherent state \cite{radcliffe,arecchi}. Note that this is analogous to the definition of the Q-function in position/momentum space, the latter given by Eq.\ \eqref{husimi}. As for the Husimi or Q-function in phase space, this distribution does not give the correct marginal distributions. This means that integrating over the $\varphi$ angle does not leave the correct distribution function for the $\theta$ variable. However, it still contains all the information about the system and it can be used to calculate the expectation value of any observable, just like in the density matrix formalism. The mapping between spin operators and the corresponding spin-space functions will be considered in detail in Section \ref{sec:2}. The main reason for choosing to work with this particular distribution function for the spin variable is that it is a function on the unit sphere and hence resemble the classical picture of a dipole moment, in fact, the evolution equation in the long scale length limit (see Eq.\ \eqref{eq:semi-classical}) is almost identical to an equation derived previously from a semiclassical treatment of the spin \cite{newarticle}. The Q-function for the spin is also nonnegative which may be desirable in some cases.

\section{The density matrix description} 
\label{sec:1}
In order to derive our phase space model we here start from the density matrix description for a spin-1/2 particle. The basis states we will use are $\left| \mathbf x, \alpha \right> = \left| \mathbf x \right> \otimes \left| \alpha \right>$ where $\left| \mathbf x \right>$ is the state with position definitely at position $\mathbf x$ and $\left| \alpha \right>$ is the state with spin up ($\alpha =1$) and spin down ($\alpha = 2$) along the axis of quantization, which we here take to be in the $z$-direction. The density matrix in this basis is then
\begin{equation} 
	\rho (\mathbf x, \alpha; \mathbf y, \beta, t) \equiv 
	\left< \mathbf x, \alpha \right| \hat \rho \left| \mathbf y, \beta \right> =
	\sum_i p_i \psi_{i}(\mathbf x,\alpha,t) \psi^*_{i} (\mathbf y,\beta,t) ,
\end{equation}
where $p_i$ is the probability to have state $\psi_i$ and the greek letters denotes the spin indexes. Here $\psi(\mathbf x, 1)$ and $\psi(\mathbf x,2)$ gives, respectively, the probability amplitude to have spin up and spin down.

The Hamiltonian for a particle in an external electromagnetic field is given by
\begin{equation}
	\hat H = 
	\frac{1}{2m} 
	\left[ \hat{\mathbf p} - q \mathbf A(\hat{\mathbf x} ,t ) \right]^2  
	+ q V (\hat{\mathbf x},t) 
	- \mu \mathbf B(\hat{\mathbf x},t) \cdot \bm \sigma ,
	\label{hamiltonian}
\end{equation}
where $m$ is the mass of the particle, $q$ is the charge (for an electron $q=-e<0$ where $e$ is the elementary charge), $\mu$ is the magnetic moment of the particle, which for electrons is given by the (signed) Bohr magneton $\mu_e = - e \hbar /(2m_e)$, $\mathbf A$ and $V$ are the electromagnetic potentials, and $\mathbf B = \nabla_x \times \mathbf A$ is the magnetic field. $\bm \sigma$ is the vector containing the three Pauli matrices as its components. With the axis of quantization in the $z$-direction they are given by 
\begin{equation} \nonumber
\sigma^{(x)} = \left( \begin{array}{cc} 0 & 1 \\ 1 & 0 \end{array} \right), \, \,
\sigma^{(y)} = \left( \begin{array}{cc} 0 & - i \\i & 0 \end{array} \right) 
\,\, \mathrm{and} \,\, 
\sigma^{(z)} = \left( \begin{array}{cc} 1 & 0 \\ 0 & -1\end{array} \right). 
\end{equation}
We will use the notation $\bm \sigma (\alpha, \beta) \equiv (\sigma^{(x)}(\alpha,\beta),\sigma^{(y)}(\alpha,\beta),\sigma^{(z)}(\alpha,\beta))$, where $\sigma^{(x)}(\alpha,\beta)$ denotes the component on row $\alpha$ and column $\beta$ of $\sigma^{(x)}$ and similarly for the $\sigma^{(y)}$ and $\sigma^{(z)}$ matrices.

The evolution equation for the density matrix can be derived from the Schr\"odinger equation for the wave function and its complex conjugate, giving the von Neumann equation  
\begin{equation}
	i \hbar \frac{\partial \hat \rho}{\partial t} = \left[ \hat H , \hat \rho \right].
	\label{vonneumann}
\end{equation}
Using the basis described above and the Hamiltonian \eqref{hamiltonian} we obtain
\begin{equation}
\begin{split}
	&
	i\hbar \frac{\partial \rho
	(\mathbf x, \alpha; \mathbf y, \beta,t)}{\partial t}  
	=
	- \frac{\hbar^2}{2m} 
		\left[ \nabla_{x}^2 - \nabla_{y}^2 \right] 
		\rho(\mathbf x, \alpha; \mathbf y,\beta,t) 
	+ \frac{i\hbar q}{m} \left[ 
		\mathbf A(\mathbf x,t) \cdot \nabla_{x}
		+ \mathbf A(\mathbf y,t) \cdot \nabla_{y}
		\right] \rho(\mathbf x, \alpha; \mathbf y,\beta,t)
	\\ & \quad
	+ \frac{q^2}{2m} \left[ 
			A^2(\mathbf x,t) - A^2(\mathbf y,t)
				\right] \rho_{\alpha\beta}(\mathbf x,\mathbf y,t)
	+ q \left[ V(\mathbf x,t) - V(\mathbf y,t) \right] 
	\rho(\mathbf x, \alpha; \mathbf y,\beta,t)
	\\ & \quad
	- \mu\sum_{\gamma =1}^2 \left[
	\mathbf B(\mathbf x,t) \cdot \bm \sigma(\alpha, \gamma) 
	\rho(\mathbf x, \gamma; \mathbf y,\beta,t)
	-
	\mathbf B(\mathbf y,t) \cdot \bm \sigma^*(\beta, \gamma)   
	\rho(\mathbf x, \alpha; \mathbf y,\gamma,t)\right],
	\label{vonnuemann}
\end{split}
\end{equation}
where we have used the Coulomb gauge $\nabla_{x} \cdot \mathbf A = 0$. In general, the evolution equation of the diagonal terms $\rho(\mathbf x, \alpha; \mathbf y, \alpha)$, $\alpha = 1,2$ are coupled via the off-diagonal terms. However, for static fields it is possible to obtain two decoupled equations for the diagonal elements, by orienting the axes so that the magnetic field is in the direction of the axis of quantization \cite{arnold}. 

\section{The Wigner and Q transformation} \label{sec:2}

The Wigner transformation for a spin-$1/2$ particle is given by 
\begin{equation}
	W(\mathbf x, \mathbf p, \alpha, \beta) = 
	\frac{1}{(2\pi\hbar)^3} \int d^3 z
	e^{-i\mathbf p \cdot \mathbf z/\hbar} 
	\rho\left(\mathbf x + {\mathbf z}/{2} , \alpha;  
	\mathbf x - {\mathbf z}/{2}, \beta \right)
	\label{4}
\end{equation}
where we have emphasized that for a particle with spin the Wigner transform must be taken for each spin matrix element of the density matrix separately. The Wigner transform of the spin density matrix has been calculated previously \cite{carruthers,oconnell,arnold}. One approach is to consider the different components of the Wigner matrix $W(\mathbf x, \mathbf p, \alpha, \beta)$, for $\alpha = 1,2$ and $\beta = 1,2$ and derive evolution equations for $W(\mathbf x, \mathbf p , 1, 1)$ and $W(\mathbf x, \mathbf p, 2,2)$ which, as the for the density matrix, are in general coupled via the off-diagonal terms \cite{arnold,manfredi2008}. Another approach is to define a quasi-distribution function for the spin degree of freedom. This can, as have been discussed above, be done in a variety of different ways \cite{cohen,chandler,scully}. The way which is a direct generalization of the Wigner function is to consider two different spin components in two arbitrary directions $s_1$ and $s_2$, corresponding to the two operators $\bm \sigma_1$ and $\bm \sigma_2$, see Ref.\ \cite{cohen}. Since the two operators in general do not commute, the values of $s_1$ and $s_2$ cannot be known simultaneously. This manifests itself in that the Wigner function $W(s_1,s_2)$, as for the corresponding case of position and momentum, can take on negative values. Another possible choice of distribution function (corresponding to anti-normal operator ordering) is the Q-function. In the position/momentum space, this distribution function is the Gaussian averaged Wigner function and due to this it is positive definite. In optics, the Q-function can be measured directly \cite{Leonhardt1997}. The spin Q-function \cite{scully} gives the probability to measure the spin in a given direction and it is this we will here use to describe the spin degree of freedom.

To derive an evolution equation for the extended phase space distribution function $f(\mathbf r,\mathbf p, \hat{\mathbf{s}})$, where $\hat{ \mathbf{s} }$ is a unit vector (not an operator), we impose the following properties:
\begin{equation}
	f(\mathbf x, \hat{ \mathbf{s} } ) = \int d^3 p\, f(\mathbf x, \mathbf p,
	\hat{\mathbf{s}}) ,
	\label{propa}
\end{equation}
should give the probability density to find the particle at position $\mathbf r$ with spin up in the direction of $\hat{ \mathbf{s} }$ and, similarly, 
\begin{equation}
	f(\mathbf p, \hat{ \mathbf{s} } ) = \int d^3 x\, f(\mathbf x,\mathbf p,
	\hat{ \mathbf{s} }) 
	\label{propb}
\end{equation}
should give the probability to have momentum $\mathbf p$ and spin up in the $\hat{ \mathbf{s} }$ direction, a direct extension of the marginal distribution conditions (\ref{eq:marginal_x}) and (\ref{eq:marginal_p}). In order to derive the distribution function in the extended phase space we note that for a state $\psi(\mathbf x, \alpha)$ we have the probabilities $|\psi(\mathbf x, 1)|^2$ ($|\psi(\mathbf x, 2)|^2$) to measure spin up (spin down) in the $z$-direction. The corresponding density matrix is given by $\rho(\mathbf x, \alpha; \mathbf y, \beta) = \psi(\mathbf x, \alpha) \psi^*(\mathbf y, \beta)$. We can then write the probability to measure spin-up in the direction of the unit vector $\hat{ \mathbf{s} }$ as 
\begin{equation}
\textrm{Tr} ( \hat P_{\uparrow}(\hat{ \mathbf{s} }) \rho ) = 
	\sum_{\alpha,\beta=1}^2 \frac{1}{2} \left[ \delta_{\alpha \beta} 
	+ \hat{ \mathbf{s} } \cdot \bm \sigma(\alpha, \beta) \right]
	\rho(\mathbf x, \beta; \mathbf x, \alpha) , 
\end{equation}
where we have defined the  (Hermitian) operator
\begin{equation}
	\hat P_{\uparrow} (\hat{ \mathbf{s} }) = \frac{1}{2} \left[ 
	1 + \hat{ \mathbf{s} } \cdot \bm \sigma \right] 
	\label{7}
\end{equation}
and where $\delta_{\alpha\beta}$ denotes the Kronecker delta. 
As an example we consider the the probability to measure the spin in the direction $\hat{ \mathbf{s} }=- \hat{\mathbf z}$ and we get get
\begin{equation}
	\sum_{\alpha,\beta=1}^2 \frac{1}{2} \left[ \delta_{\alpha\beta} - \sigma^{(z)}(\alpha, \beta) \right]
	\rho(\mathbf x, \beta; \mathbf x, \alpha) = |\psi_2(\mathbf x)|^2 ,
\end{equation}
as we expect (note that measuring spin-up in the $-\bm{\mathbf z}$ direction is equivalent to measure spin-down in the $\bm{\mathbf z}$ direction). 
The generalization to a statistical distribution of states is straightforward. Using the Wigner transform for the position and momentum and the spin transform discussed above we obtain the function
\begin{equation}
	f(\mathbf x,\mathbf p,\hat{ \mathbf{s} }) = 
	\sum_{\alpha,\beta=1}^2 
	\frac{1}{2} 
	\left[ \delta_{\alpha \beta} + \hat{ \mathbf{s} } \cdot \bm \sigma(\alpha, \beta)
	\right]
	W (\mathbf x, \mathbf p, \beta, \alpha) 
\end{equation}
which have the properties \eqref{propa} and \eqref{propb} stated above. 
The function $f$ may also be written as
\begin{equation}
	f(\mathbf x, \mathbf p, \hat{ \mathbf{s} }) = 
	\mathrm{Tr} 
	\left[ \hat P_{\uparrow} (\hat{ \mathbf{s} }) 
	\overline{W} (\mathbf x, \mathbf p) \right] ,
	\label{9}
\end{equation}
where $\overline W$ is the $2 \times 2$ matrix with elements $W(\mathbf x, \mathbf p, \alpha, \beta)$. 

The normalization of the extended Wigner function is given by
\begin{equation}
	\mathrm{Tr} \int d^3 x\, d^3 p\, d^2\hat{s }\, 
	\frac{1}{2} \left( 1 + \hat{ \mathbf{s} }\cdot \bm \sigma\right) W = 
	2\pi .
\end{equation}
Hence we obtain a distribution function which is normalized over the allowed spin values if we redefine the operator in Eq.\ \eqref{7} as 
\begin{equation}
	\hat P_\uparrow (\hat{ \mathbf{s} }) \equiv \frac{1}{4\pi} \left( 1 + \hat{ \mathbf{s} }\cdot \bm 
	\sigma \right) .
	\label{11}
\end{equation}
In the Wigner formalism without spin, the density matrix is transformed into the Wigner function and the operators are transformed into phase space functions. For an operator $\hat{g}= g(\hat{\mathbf x}, \hat{\mathbf p})$ the corresponding phase space function is given by
\begin{equation}
	g(\mathbf x, \mathbf p) = 
	\int d^3 z e^{-\frac{i}{\hbar} \mathbf p \cdot \mathbf z}
	\left< \mathbf x + \frac{\mathbf z}{2} \right| 
	\hat g \left| \mathbf x - \frac{\mathbf z}{2} \right>. 
	\label{13}
\end{equation}
It can also be obtained by using Weyl ordering as described in subsection 2.3. 
With this function the expectation value of the operator is calculated as a phase space integral 
\begin{equation}
	\left< \hat{g} \right> = 
	\int d^3 x\, d^3 p\, f(\mathbf x, \mathbf p) g(\mathbf x, \mathbf p) = \textrm{Tr} (\hat \rho \hat{g}) .
\end{equation}
where $f(\mathbf x, \mathbf p)$ and $\hat \rho$ are related via a Wigner transform. In analogy with this, for a given operator $\hat h$ acting on the spin degree of freedom, we define the corresponding spin-space function
\begin{eqnarray}
	h(\hat{ \mathbf{s} }) = \mathrm{Tr} \left[  
	\frac{1}{2} \left( 1 + 3 \hat{ \mathbf{s} } \cdot \bm \sigma \right) \hat h \right]
	= 
	\sum_{\alpha, \beta = 1}^2 \frac{1}{2} \left[\delta_{\alpha \beta} + 
	3 \hat{ \mathbf{s} }\cdot \bm \sigma(\alpha, \beta)  \right] h(\beta, \alpha) ,
	\label{15}
\end{eqnarray}
where $h(\alpha, \beta)$ denotes the $(\alpha, \beta)$ component of the operator $\hat h$. 
With this definition the expectation value of the operator is now calculated as an integral over the possible spin directions according to
\begin{eqnarray}
	\textrm{Tr} (\hat \rho \hat h) = \int d^2\hat{s} f(\hat{ \mathbf{s} }) h(\hat{ \mathbf{s} }) 
\end{eqnarray}
where we have used $\int d^2\hat{s}\, s_a s_b = (4\pi/3 )\delta_{a b}$ and $\sigma^i_{\alpha \beta} \sigma^i_{\gamma \delta} = 2 \delta_{\alpha \delta} \delta_{\beta \gamma} - \delta_{\alpha \beta} \delta_{\gamma \delta}$. In doing the calculation above we have also used that the general form of a spin-operator is $\hat h = a I  + \mathbf b \cdot \bm \sigma$ where $a$ and $\mathbf b$ may be dependent of position and momenta. Note that the definition of the spin space function, Eq.\ \eqref{15} implies that the spin operator $\bm \sigma$ is related to the spin unit vector $\hat{\mathbf{s}}$ according to 
\begin{equation}
	\bm \sigma \rightarrow 3 \hat{ \mathbf{s} }. 
\end{equation}
For operators depending on both the position and momentum and the spin degree of freedom the corresponding extended phase space function is obtained by doing both the transformations \eqref{13} and \eqref{15}.

The operator \eqref{11} can also be written 
$\hat P (\hat{ \mathbf{s} }) = \left| \hat{ \mathbf{s} } \right> \left< \hat{ \mathbf{s} }\right|$, where $\left| \hat{ \mathbf{s} } \right>$ is the spin coherent state \cite{radcliffe,arecchi}. The definition \eqref{9} is then seen to coincide with the definition of the spin Q-function, see Eq.\ \eqref{q-spin}. The function $f(\mathbf x, \mathbf p, \hat{ \mathbf{s} })$ is hence a combination of a Wigner function in the phase space variables and the Q-function for the spin. 

\subsection{Equivalence with the density matrix formalism}

The construction above contains the same information as the density matrix, and the distribution function can be used to calculate the expectation value of any observable. A more direct way to see the equivalence is to note that, for a given distribution function $f(\mathbf x, \mathbf p, \hat{\mathbf s})$, it is possible to obtain the corresponding Wigner matrix as
\begin{equation}
	\overline W(\mathbf x, \mathbf p)  
	= \left( \begin{array}{cc} \rho_{11} & \rho_{12} \\ \rho_{21} & \rho_{22} 
	\end{array} \right) = 
	\int d^2 \hat s f(\mathbf x ,\mathbf p , \hat{\mathbf s}) \frac{1}{2}
	\left( \begin{array}{cc} 1 + 3s_z & 3( s_x - i s_y) \\
	3 (s_x + i s_y ) & 1 - 3 s_z \end{array} \right). 
\end{equation}
From this it is the possible to obtain the density matrix by taking the inverse Wigner transform, (see for example Ref.\ \cite{mizrahi}).

\section{Evolution equation}
\label{sec:evo}

To derive the evolution equation for $f(\mathbf{x}, \mathbf{p}, \hat{\mathbf{s}})$, the Wigner transform of Eq.\ \eqref{vonnuemann} is calculated with the result (assuming that the fields and potentials are analytic functions)
\begin{equation}
\begin{split}
	&
	\left(\frac{\partial }{\partial t} 
	+ \frac{1}{m} \mathbf p \cdot \nabla_{x}\right)
	W(\mathbf x, \mathbf p, \alpha, \beta)
	= \frac{q}{m}\Bigg[
	 \mathbf A(\mathbf x) 
	\cdot \overrightarrow\nabla_{x}
	\cos \left( \frac{\hbar}{2} 
	\overleftarrow\nabla_{x}
	\cdot \overrightarrow\nabla_{p}
	\right)
\\ & \qquad
	   - \frac{2}{\hbar} \mathbf p \cdot \mathbf A(\mathbf x) 
	\sin \left( \frac{\hbar}{2} 
	\overleftarrow\nabla_{x}
	\cdot \overrightarrow\nabla_{p} \right)
	+ \frac{q}{\hbar} A^2(\mathbf x) 
	\sin \left(\frac{\hbar}{2} 
	\overleftarrow \nabla_{x} \cdot
	\overrightarrow \nabla_{p} \right) \Bigg]
	W(\mathbf x, \mathbf p, \alpha, \beta) 
		+ \frac{2 q}{\hbar} V(\mathbf x)
	\sin \left(\frac{\hbar}{2} 
	\overleftarrow \nabla_{x} \cdot
	\overrightarrow \nabla_{p} \right) 
	W(\mathbf x, \mathbf p, \alpha, \beta)
\\ &	\qquad
	+ \frac{i\mu}{\hbar}
	\sum_{\gamma=1}^2
	\mathbf B(\mathbf x)  \cdot  \Bigg[
	\bm \sigma(\alpha, \gamma) 
	\exp\left(\frac{i\hbar}{2} \overleftarrow \nabla_{x}
	\cdot \overrightarrow \nabla_{p} \right)
	W(\mathbf x, \mathbf p, \gamma, \beta)
	-  \bm \sigma^*(\beta, \gamma)
	\exp\left(-\frac{i\hbar}{2} \overleftarrow \nabla_{x}
	\cdot \overrightarrow \nabla_{p} \right)
	W(\mathbf x, \mathbf p, \alpha, \gamma) \Bigg],
\end{split}
\end{equation}
where functions of an operator is defined by its formal Taylor expansion and the left (right) arrow above the differential operators indicate that they act on the functions on the left (right). If the potentials have discontinuities the above equation can instead be written explicitly in the form of an integro-differential equation. Next we multiply by $[\delta_{\beta\alpha} + \hat{ \mathbf{s} } \cdot \bm \sigma(\beta,\alpha) ]/2$ and sum over $\alpha$ and $\beta$. 
The operators acting on the left hand side and the first four terms on the right hand side commute with the Pauli matrices and for these we obtain $W(\mathbf x, \mathbf p, \alpha, \beta) \rightarrow f(\mathbf x, \mathbf p, \hat{ \mathbf{s} })$. For the last two terms we use the property 
\begin{equation}
	\sum_{\gamma = 1}^2
	\mathbf A \cdot \bm \sigma(\alpha, \gamma) 
	\mathbf B \cdot \bm \sigma(\gamma, \beta) =  
	\mathbf A \cdot \mathbf B \delta_{\alpha \beta} 
	+ i[\bm \sigma(\alpha,\beta) \cdot (\mathbf A \times
	\mathbf B)] 
\end{equation}
and also that $\bm \sigma_{\alpha \beta}^* = \bm \sigma_{\beta \alpha}$. 
After some straightforward calculations we get the evolution equation for the extended Wigner function   
\begin{equation}\label{eq:full-wigner-equation}
\begin{split}
&
	\left( \frac{\partial }{\partial t}
	+ \frac{1}{m} \mathbf p \cdot \nabla_{x} \right) f(\mathbf x, \mathbf p, \hat{ \mathbf{s} }) 
	= \Bigg\{\bigg[
	\left( 
	- \frac{q}{m} \mathbf p \cdot \mathbf A
	+ \frac{q^2}{2m} A^2 
	+ q V
	\right) 
	- \mu \left( \mathbf B \cdot \overrightarrow \nabla_{\hat{ \mathbf{s} }} 
	+ \hat{ {s} } \cdot \mathbf B 
	\right)
	\bigg]
	\frac{2}{\hbar}
	\sin \left( \frac{\hbar}{2} 
	\overleftarrow \nabla_{x} \cdot 
	\overrightarrow \nabla_{p} \right) 
\\ &\qquad\qquad\qquad\qquad\qquad\qquad
	+\left[\frac{q}{m} \mathbf A \cdot \overrightarrow \nabla_{ x}
	- \frac{2 \mu}{\hbar} (\hat{ \mathbf{s} }\times \mathbf B)
	\cdot \overrightarrow \nabla_{\hat{ {s} }} \right]
	\cos \left( \frac{\hbar}{2} 
	\overleftarrow \nabla_{x} 
	\cdot \overrightarrow \nabla_{p}\right)
	\Bigg\}
	f(\mathbf x,\mathbf p,\hat{ \mathbf{s} }).
\end{split}
\end{equation}
An advantage of writing the evolution equation in this form is that we may Taylor expand the trigonometric function to sufficient order in $\hbar$ to obtain the semi-classical limit directly. 

Next we make a variable transformation in the evolution equation. 
The canonical momentum $\mathbf p$ is related to the velocity by $\mathbf v = (\mathbf p - q \mathbf A)/m$. Changing variables from $\mathbf x$, $\mathbf p$ and $t$ to $\mathbf x$, $\mathbf v$, $t$ 
we get
\begin{subequations} \label{variabelsubs}
\begin{eqnarray}
	\nabla_{x i} 
	&\rightarrow & \nabla_{x i} - \frac{q}{m}  \sum_{j=1}^3( \nabla_{x i} 
	A_j) \nabla_{v j} 
	\label{17a} \\ 
	\nabla_{p i} &\rightarrow & 
	\frac{1}{m} \nabla_{v i} 
\end{eqnarray}
where $\nabla_{x i} = \partial/\partial x_i$ and $\nabla_{v i} = \partial/\partial v_i$. For the time derivative we get
\begin{equation} \label{timederivative}
	\partial_t \rightarrow \partial_t - \frac{q}{m} \sum_{i=1}^3  [\partial_t A_i(\mathbf x)] \nabla_{v i} . 
\end{equation}
\end{subequations}
We can then write the full quantum-kinetic equation (\ref{eq:full-wigner-equation}) as
\begin{equation} 
\label{eq:full-wigner-equation-2}
\begin{split}
	&
	\frac{\partial f}{\partial t} + \mathbf v \cdot \nabla_{x} f 
	+ \left[
		\frac{q}{m}(\mathbf{E} + \mathbf{v}\times\mathbf{B})
		+ \frac{\mu}{m}\nabla_x[(\overrightarrow \nabla_{\hat{ {s} }} 
		+ \hat{\mathbf{s} }) \cdot \mathbf B)
	\right]\cdot\nabla_vf
	+ \frac{2 \mu}{\hbar} (\hat{ \mathbf{s} }\times \mathbf B)
		\cdot \nabla_{\hat{ {s} }} f
\\ &\quad
	= 
	\left[
		\frac{q}{m}\left(V - \mathbf{v}\cdot\mathbf{A} \right)
		- \frac{\mu}{m} \left( \mathbf B \cdot \overrightarrow \nabla_{\hat{ {s} }} 
		+ \hat{ \mathbf{s} } \cdot \mathbf B 
	\right)
	\right]
	\left[
		 \frac{2m}{\hbar} \sin\left(\frac{\hbar}{2m}\overleftarrow{\nabla}_x\cdot\overrightarrow{\nabla}_v \right)
		- \overleftarrow{\nabla}_x\cdot\overrightarrow{\nabla}_v
	\right]f
\\ & \qquad
	+ \left[\frac{q}{m} \mathbf A \cdot \overrightarrow \nabla_{x}
		- \frac{q^2}{m^2}[(\mathbf{A}\cdot\nabla_x)\mathbf{A}]\cdot\nabla_v
		- \frac{2 \mu}{\hbar} (\hat{ \mathbf{s} }\times \mathbf B)
		\cdot \overrightarrow \nabla_{\hat{ {s} }} \right]
		\left[
		\cos \left( \frac{\hbar}{2m} 
		\overleftarrow \nabla_{x} 
		\cdot \overrightarrow \nabla_{v}\right) - 1
	\right] f .
\end{split}
\end{equation}
displaying the classical and semiclassical terms more explicitly on the left-hand side of the equation. We note that the terms on the right-hand side all are higher-order derivative corrections.

\section{Many-particle evolution equation}
\label{sec:many}

So far we have only considered one particle in an external electromagnetic field. To make a straightforward generalization to an $N$-body system we will consider the mean field approximation. In order to keep things simple we will neglect effects due to spin statistics (antisymmetry of the wavefunction). To a certain degree such effects can be incorporated by choosing appropriate initial conditions, see the next section. Introducing the many-particle density matrix $\hat \rho_{1\dots N}$ it will satisfy the von Neumann equation
\begin{equation}
	i \hbar \frac{\partial \hat \rho_{1 \dots N}}{\partial t} = \left[ \hat H^{(N)} , \hat \rho_{1\dots N}\right] .
\end{equation}
The $N$-body Hamiltonian $\hat H^{(N)}$ in general includes interactions between the particles
\begin{equation}
	\hat H^{(N)} = \sum_{i = 1}^N \hat H_i + \sum_{i<j=1} ^N \hat H_{ij} ,
	\label{manyham}
\end{equation}
where $\hat H_i = (\hat{\mathbf p}_i - q \mathbf A_0(\hat{\mathbf x}_i))^2/2m + q V_0 (\hat{\mathbf x}_i)$ is the Hamiltonian for particle $i$ and contains the kinetic energy and the interaction with an external electromagnetic field $(V_0,\mathbf A_0)$, and $\hat H_{ij}$ is the interaction between particle $i$ and $j$ which we assume to be the full electromagnetic interaction between the particles. The interaction is hence obtained by solving Maxwell's equations. Following Ref.\ \cite{bonitz} we introduce the reduced density matrix in the thermodynamic limit ($N, V \rightarrow \infty$, $N/V = n_0 = \textrm{const.}$) 
\begin{equation}
	\hat \rho_{1 \dots s} = V^s \mathrm{Tr}_{s+1, \dots, N} \hat \rho_{1 \dots N}, 
\end{equation}
where $V$ is the volume of the system and the trace includes summing over the spin degree of freedom. The normalization is given by
\begin{equation}
	\frac{1}{V^s} \mathrm{Tr}_{1,\dots s} \hat \rho_{1\dots s}= 1 .
	\label{43}
\end{equation}
Note that this means that for the diagonal elements of the one-particle reduced density matrix $\rho_1(\mathbf x, \mathbf x)$ is proportional to the probability density to find any one of the $N$ particles in position $\mathbf x$ independently of the positions of all the other particles. 
Expectation values of an $s$-body operator is given by 
\begin{equation}
	\left< \hat A_{1 \dots s}  \right> = \frac{n_0^s}{s!} \mathrm{Tr} \hat A_{1\dots s} \hat \rho_{1 \dots s}
\end{equation}
The evolution equations for the reduced density matrix is given by the BBGKY-hierarchy \cite{bogolyubov} 
\begin{equation}
	i \hbar \frac{\partial \hat \rho_{1\dots s}}{\partial t} - \left[ \hat H^{(s)} , \hat \rho_{1\dots s} \right] =
	n_0 \mathrm{Tr}_{s+1} \sum_{i=1}^s \left[ \hat H_{i, s+1} , \hat \rho_{1\dots s+1} \right], 
\end{equation}
where $H^{(s)}$ is obtained by changing $N \rightarrow s$ in Eq.\ \eqref{manyham}.
Considering the first order equation and introducing the two-particle correlation as $\hat \rho_{12} = \hat \rho_{1} \hat \rho_{2} + \hat g_{12}$ we may write this as 
\begin{equation}
	i\hbar \frac{\partial \hat \rho_1}{\partial t} - \left[\hat H^{(1)} , \hat \rho_1 \right]  
	- \left[ \hat H_{\textrm{MF}} , \hat \rho_1 \right] = n_0 \mathrm{Tr}_2 \left[ 
	\hat H_{12} , \hat g_{12} \right] ,
	\label{45}
\end{equation}
where the $\hat H_{MF} = \mathrm{Tr}_2 \hat H_{12} \hat \rho_2$ is the mean field which is found by solving Maxwell's equations self-consistently. The effects of particle-particle scattering is included in the correlation operator $\hat g_{12}$. This, in turn, satisfies an equation that is coupled to the three particle correlations and so on. Here we will be mainly interested in the collective effects of the plasma and hence we will neglect the right hand side of Eq.\ \eqref{45}, i.e.\ use the Hartree approximation. In order to include self-energy effects and ionization/recombination it is necessary to keep higher order correlations. 

Comparing Eq.\ \eqref{45} with the corresponding equation for a single particle in an external electromagnetic field, Eq.\ \eqref{vonneumann}, we note that they are formally the same. Thus in order to include systems of $N$-particles in the Hartree approximation we hence need to assume that the fields in the evolution equation are the self consistent fields and then keep in mind that the density matrix is now normalized according to Eq.\ \eqref{43}. However, since we will not pursue the issue of the quantum BBGKY-hierarchy further, we may redefine the one-particle distribution function so that it has the normalization
\begin{equation}
	\mathrm Tr \hat \rho_1 = n_0 , 
\end{equation}
so that for example $\left< \mathbf x \right| \hat \rho_1 \left| \mathbf x \right> = n(\mathbf x)$ gives the mean density of particles at position $\mathbf x$. 

The mean field interaction $\hat H_{\mathrm{MF}}$ is obtained by coupling the equation to Maxwell's equations. The expression for the charge and current densities are then
\begin{eqnarray}
	n(\mathbf x,t) &=& q \int d^3 v\, d^2 \hat{s}\, f(\mathbf x, \mathbf v, \hat{\mathbf s}, t) 
	\label{chargedensity}
	\\
	\mathbf j (\mathbf x, t ) &=& \mathbf j_f(\mathbf x, t) + \mathbf j_M (\mathbf x,t) = 
	q \int d^3 v\, d^2 \hat{s}\, f(\mathbf x, \mathbf v, \hat{\mathbf s} ,t) 
	\mathbf v + \mu\nabla_x \times  \int d^3 v\, d^2 \hat{s}\, f(\mathbf x, \mathbf v, \hat{\mathbf s},t) 
	3 \hat{\mathbf s}, 
	\label{currentdensity}
\end{eqnarray}	
where we thus obtain a magnetization current contribution due to the spin (see Ref.\ \cite{suttorp}).

\section{Thermodynamic equilibrium density matrix} 

As an example we calculate the extended phase space distribution function for a system of $N$ non-interacting particles in a constant magnetic field which are in thermodynamic equilibrium at temperature $T$.  Assuming that the magnetic field is $\mathbf B = B_0 \hat{\mathbf z}$ and using the Landau gauge $\mathbf A = (-y B_0, 0 , 0)$ we can obtain the eigenstates of the Hamiltonian, Eq.\ \eqref{hamiltonian}, 
\begin{equation}
	\psi_{p_x,n,p_z,a} (x,y,z,\alpha) = \frac{e^{\frac{i}{\hbar}( p_x x + p_z z) }}{2\pi\hbar} \phi_n\left(y + \frac{p_x}{qB}\right) \chi_a(\alpha)
	\label{states}
\end{equation}
where $\phi_n$ is the $n$'th harmonic oscillator wave function given by 
\begin{equation}
	\phi_n(y) = \frac{1}{\sqrt{2^n n!}} \left( \frac{m \omega}{\pi \hbar} \right)^{1/4} 
	\exp \left( - \frac{m \omega}{2 \hbar} y^2 \right)
	H_n \left(\sqrt{\frac{m\omega}{\hbar}}  y \right) ,
\end{equation}
where $H_n$ are the Hermite polynomials \cite{handbook}. Furthermore, we have introduced the spinors $\chi_a(\alpha)$ which satisfies $\sigma_z \chi_a (\alpha) = a \chi_a (\alpha)$ with $a = \pm 1$.  The energy levels corresponding to Eq.\ \eqref{states} are given by 
\begin{equation}
	E_{n , p_z, a} = \hbar \omega_c \left( \frac{1}{2} + n \right) 
	+ \frac{p_z^2}{2m} 
	- a \mu_B B_0 .
\end{equation}
Note that the energy is independent of the momentum in the $x$-direction so that the energy levels are degenerate. 
The thermal equilibrium density matrix at temperature $T$ is given by 
\begin{equation}
	\hat \rho = \frac{e^{-\hat H / k_B T}}{Z} ,
\end{equation}
where $k_B$ is Boltzmann's constant and the partition function is given by
\begin{equation}
	Z = \mathrm{Tr}e^{-\hat H / k_B T} .
\end{equation} 
Considering the one-particle density matrix, it can be written
\begin{equation}
	\rho(\mathbf x, \alpha; \mathbf y, \beta) = \sum_{p_x, n, p_z, a} p_{p_x, n, p_z, a}
	\psi_{p_x, n, p_z, a}(\mathbf x, \alpha) 
	\psi^*_{p_x, n, p_z, a}(\mathbf y,\beta) 
	\chi_a (\alpha) \chi^\dagger _a(\beta) ,
\end{equation}
where the probability for the state with quantum numbers $(p_x, n, p_y, a)$ is given by
\begin{equation}
	p_{p_x, n ,p_z, a} = \frac{1}{e^{(E_{n,p_z,a}- \mu_c)/k_B T} +1} ,
\end{equation}
where $\mu_c$ is the chemical potential. The Wigner transform of the density matrix for an harmonic oscillator has been calculated by Ref.\ \cite{Schleich2001}. Using their result we can calculate the Wigner transform to be 
\begin{equation}
\begin{split}
	&
    	W(\mathbf x, \mathbf p, \alpha, \beta) = \sum_{n, a} 
    	\frac{1}{e^{(E_{n,p_z,a} - \mu_c)/T} + 1} \frac{2(-1)^n}{(2\pi\hbar)^3} 
	\exp \left[ -\frac{2}{\hbar\omega} 
	\left( \frac{p_y^2}{2m} + \frac{m\omega^2}{2} 
	\left(y + \frac{p_x}{qB}\right)^2
	\right)\right] 
\\ & \qquad \times
	L_n \left[ \frac{4}{\hbar\omega} \left( \frac{p_y^2}{2m} + \frac{m\omega^2 }{2} \left(y + \frac{p_x}{qB}\right)^2 \right)\right] 
	\chi_{a} (\alpha) \chi_{a} (\alpha'),
\end{split}
\end{equation}
where $L_n$ denotes the Laguerre polynomials \cite{handbook}. Calculating the spin-transform, Eq.\ \eqref{9}, of this and also changing variables to $\mathbf v = (\mathbf p - q \mathbf A)/m$ we finally obtain 
\begin{equation}
\begin{split}
	&
	f(\mathbf x, \mathbf v,\hat{ \mathbf{s} }) = 
	\sum_{n,a} \frac{n_0 (-1)^n}{2 \pi (2\pi\hbar)^3} 
	   \frac{ 1 + a \cos\theta_s }{e^{(E_{n,p_z,a} - \mu_c)/T} + 1}  
	\exp \left[ -\frac{2}{\hbar\omega} 
	\left( \frac{m (v_x^2 + v_y^2)}{2}
	\right)\right] 
	   L_n \left[ \frac{4}{\hbar\omega} \left( \frac{m(v_x^2 + v_y^2)}{2} \right)\right] ,
\label{eq:equilibrium}
\end{split}
\end{equation}
where we have also multiplied by $n_0$ to obtain the chosen normalization (see the previous section).
Note that the argument appearing in the exponential and the Laguerre polynomials is just the kinetic energy of the motion perpendicular to the magnetic field. However, as opposed to the classical case, $v_x$ and $v_y$ are non-commuting quantities and cannot be determined simultaneously. To verify that Eq.\ \eqref{eq:equilibrium} indeed is a solution to the Wigner equation we note that for stationary solutions in the given choice of magnetic field the equation can be written 
\begin{equation}
	( v_x \partial_x + v_y \partial_y ) f = 0 ,
\end{equation}
and we see that in fact any spatially homogenous function solves this. 

The expression \eqref{eq:equilibrium} contains Landau-quantization, spin splitting of energy states and Fermi-Dirac statistics. For cases where the chemical potential $\mu_c$ is large, and the difference between nearby Landau levels is smaller than the thermal energy, the velocity distribution approaches the classical Maxwellian. An important quantum mechanical result that remains in this limit is that the probability distribution of the spin up and down populations scales as $1+ \cos \theta_s$ and $1 - \cos \theta_s$, respectively. Thus, in the above regime the distribution can be approximated by 
\begin{equation}
	f(\mathbf x, \mathbf v, \hat{\mathbf s}) = F_+ (\mathbf v) (1+ \cos \theta_s) + 
	F_- (\mathbf v) (1 - \cos \theta_s) ,
	\label{61}
\end{equation}
where $F_\pm$ are Maxwellian distributions. The ratio $F_+/F_-$ in thermodynamic equilibrium is $F_+ /F_- = \exp(- 2 \mu_B B_0)/k_B T$. For small chemical potential, when Fermi-Dirac statistics applies to $F_\pm$, we can still have the form \eqref{61}, but the ratio $F_+/F_-$ is velocity dependent. Actually, even in the absence of thermodynamical equilibrium Eq.\ \eqref{61}  is the most general time independent, homogenous expression for the distribution function in a constant magnetic field.

\section{Long scale length limit} \label{sec:3}

To obtain the long scale limit we Taylor expand the trigonometric operators to order $\hbar$, which applies if the characteristic scale lenghts are longer than the thermal de Broglie length. Thus we henceforth neglect higher order terms in $\hbar$, such as $\frac{\hbar^2}{4} \nabla^2_{\mathbf x} V(\mathbf x) \cdot \nabla^2_{\mathbf p} f $. The evolution equation then becomes
\begin{equation}
\begin{split}
	&
	\left(\frac{\partial }{\partial t} + \frac{1}{m} \mathbf p \cdot \nabla_{x}\right) f 
	= 
	\left( -\frac{q}{m} p_i \nabla_{x j} A_i + 
	\frac{q^2}{2m} \nabla_{x j} A^2 
	+ q \nabla_{x j} V \right) \cdot
	\nabla_{p j} f 
\\ &\qquad
	- \mu
	\left( \nabla_{x j} B_i \nabla_{\hat{{s} } i} 
	+ s_i \nabla_{x j} B_i 
	\right) \cdot \nabla_{p j} f
	+ \left[ 
	\frac{q}{m} \mathbf A \cdot \nabla_{x} 
	-	\frac{2\mu}{\hbar} (\hat{ \mathbf{s} } \times \mathbf B) 
	\cdot \nabla_{\hat{ {s} }}
	\right] f .
	\label{16}
\end{split}
\end{equation}
Making a variable change from $\mathbf x$, $\mathbf p$ and $t$ to $\mathbf x$, $\mathbf v$, $t$ according to Eqs.\ \eqref{variabelsubs} the second term in Eq.\ \eqref{17a} above will combine with other terms in Eq.\ \eqref{16} to produce the magnetic field term in the Lorentz force.
The last term in the time derivative Eq.\ \eqref{timederivative} will combine with the gradient of the scalar potential $\nabla V$ to produce the electric field $\mathbf E = -\nabla V + \partial_t \mathbf A$. The evolution equation then takes the form
\begin{equation}
\begin{split}
	&
	\frac{\partial f}{\partial t} 
	+ \mathbf v \cdot \nabla_{x} f
	+  \left[
	\frac{q}{m} \left( \mathbf E + 	\mathbf v \times \mathbf B \right)
	+ \frac{\mu}{m} \nabla_{x} (\hat{ \mathbf{s} } \cdot \mathbf B) 
	\right] \cdot \nabla_{v} f
	+ \frac{2 \mu}{\hbar} (\hat{ \mathbf{s} } \times \mathbf B) 
	\cdot \nabla_{\hat{ {s} }} f 
	+ \frac{\mu}{m} 
	[\nabla_{ x} (\mathbf B \cdot \nabla_{\hat{s}})]
	\cdot \nabla_{v} f 	= 0.
\label{eq:semi-classical}
\end{split}
\end{equation}
Note that the last term contains derivatives both with respect to the velocity $\mathbf v$ and the spin $\hat{\mathbf{s}}$. The equation above has already been studied in \cite{newarticle} with the last term missing due to semi-classical approximations. It is there shown to give rise to new oscillation modes due to the anomalous magnetic moment of the electron. A similar equation to has also been studied in \cite{cowley} where it is investigated whether spin may be of importance in magnetic confined fusion experiments. The difference between Eq.\ \eqref{eq:semi-classical} and the semi-classical case (with the last term missing) is due to the fact that the quantum mechanical probability distribution is always spread out, as follows from Eq.\ \eqref{11}. In order to demonstrate this we consider the distribution function for a single particle, that at a time $t$ has a given spin state, pointing in the direction $\hat{\mathbf{e}}$, where $\hat{\mathbf{e}}$ is a unit vector. As follows from Eq.\ \eqref{11}, the corresponding distribution function, which is smeared out in spin space, can be written $f(\mathbf{x},\mathbf{v} , \hat{\mathbf{s}})=F (\mathbf{x},\mathbf{v})(1 + \hat{\mathbf{e}} \cdot\hat{\mathbf{s}})/4\pi$. If we average over all spin directions, the last term combine with the magnetic dipole term according to
\begin{eqnarray}
	\frac{1}{4\pi } \frac{\mu }{m}\int d^{2}s \nabla _{x} \left( \mathbf{B} 
	\cdot \hat{\mathbf{s}} + \mathbf{B} \cdot \nabla _{\hat s} \right) \cdot 
	\nabla_{v} F (\mathbf{x},\mathbf{v}) (1+\hat{\mathbf{e}} \cdot \hat{\mathbf{s}}) 
	= \frac{\mu}{m} \nabla _{x} \cdot 
	(\mathbf{B} \cdot \hat{\mathbf{e}}) \nabla _{v} 
	F(\mathbf{x},\mathbf{v}) 
	\label{64}
\end{eqnarray} 
where we stress that $2/3$ of the contribution comes from the latter term. The full evolution equation for this reduced distribution function can be written
\begin{equation}
\begin{split}
	&
	\frac{\partial F}{\partial t} 
	+ \mathbf v \cdot \nabla_{x} F
	+  \left[
	\frac{q}{m} \left( \mathbf E + \mathbf v \times \mathbf B \right)
	+ \frac{\mu}{m} \nabla_{x} B 
	\right] \cdot \nabla_{v} F  	= 0.
	\label{65}
\end{split}
\end{equation}
where $B = |\mathbf B|$ and we have allowed the spin direction $\hat{\mathbf e}$ to be slowly variying, following the variations of the magnetic field direction, i.e. $\hat{\mathbf e} = \hat{\mathbf B}(\mathbf x,t)$. The equation above is usefull when the spin state of each particle is conserved for a sufficiently long time.

For a semi-classical treatment of the spin we would expect that the probability to measure the spin in the direction $\hat{ \mathbf{s} }$ given that the spin is in $\hat{\mathbf e}$-direction is given by $f_{\textrm{cl}}(\hat{ \mathbf{s} }) = \delta(\hat{ \mathbf{s} } - \hat{\mathbf r})$. However, as can be seen from the above equation, the classical limit of a particle with spin in the $\hat{\mathbf r}$-direction is not a particle with definite magnetic moment in the $\hat{\mathbf r}$-direction but a statistical distribution of spins in all directions (except $\hat{\mathbf s} = -\hat{\mathbf r}$ which has zero probability). For a magnetized electron plasma, the magnetization is given by $\nabla \times \mathbf M = \nabla \times \mu \left\langle  \bm \sigma \right\rangle$ where the expectation value is taken with respect to the spin degree of freedom. In the quantum model developed here the spin is given by 
\begin{equation} 
	\left\langle \bm \sigma \right\rangle = \int d^3v\,d^2\hat{s}\, 3 \hat{ s}\, f(\mathbf x, \mathbf v, \hat{\mathbf s},t) . 
\end{equation}
The factor 3 will account for the fact that probability to find the spin in a certain direction is smeared out over the whole unit sphere. In a classical treatment of the spin variable the corresponding integral contains no factor 3, but instead the distribution function is a delta function of the spin and hence the same result can be obtained. The latter is the model used in Ref.\ \cite{newarticle}.

\section{Examples in linearized theory} \label{sec:lin}

\subsection{Spin induced damping of Alfv\'{e}n waves}
As an example of the usefulness of Eq. \eqref{eq:semi-classical} we will consider
shear Alfv\'{e}n like waves in the linear limit. First we divide the
variables as $f=f_{0}+f_{1}$ and $\mathbf{B}=\mathbf{B}_{0} + \mathbf{B}_{1}$, with in which case the linearized electron equation can be written 
\begin{equation}
\begin{split}
&
	\left[\frac{\partial}{\partial t}+\mathbf{v\cdot } \nabla_{ x}  +\frac{q_{e}}{m_{e}}%
		\left( \mathbf{v} \times \mathbf{B}_{0}\right) \cdot \nabla_{v}
		+\frac{2\mu _{e}}{ \hbar }\left( \hat{\mathbf{s}} \times \mathbf{B}_{0}\right) \cdot 
		\nabla_{\hat{{s}}}
	\right] f_{1}
\\ & \qquad
	=-\left\{ \frac{q_{e}}{m_{e}}\left( \mathbf{E} + \mathbf {v}\times \mathbf{B}_{1}\right) 
	+ \frac{\mu _{e}}{m_{e}}
	\nabla_{x} \left( \hat{\mathbf{s}} \cdot \mathbf{B}_{1}\right) 
	+\frac{\mu _{e} }{m_e}\nabla_{x} \left[ \left( \mathbf{B}
	_{1}\cdot \nabla_{\hat{ {s} }} \right) \right] \right\} \cdot 
	\nabla_{v} f_{0}
	-\frac{2\mu _{e}}{\hbar }\left( \hat{\mathbf{s}} \times \mathbf{B}_{1}\right) \nabla_{\hat{\mathbf{s}}} f_{0} 
\label{lin-long}
\end{split}
\end{equation}
The magnetic moment is given by $\mu_{e} = - (g/2) e \hbar/(2m_e)$ where we explicitly have introduced the Land\'e $g$-factor which is exactly 2 within Dirac theory, but from QED we obtain $g/2-1\simeq 0.0016$. 
The term $(\mathbf{v}\times \mathbf{B}_{1})\cdot \nabla_{v} f_{0}$ can be dropped for an isotropic equilibrium distribution, which will be used below. Furthermore, letting $\mathbf{B}_{0}=B_{0}\hat{\mathbf{z}}$, introducing cylindrical coordinates in velocity space ($v_{\bot }$, $\varphi _{v}$, $v_{z}$) and spherical
coordinates in spin space ($\varphi _{s}$, $\theta _{s}$), and making a
plane wave ansatz $f_{1}=\tilde{f}_{1}\exp [i(\mathbf{k\cdot r}-\omega t)]$, the equation is written 
\begin{equation}
\begin{split}
	\left[ i \left( \omega - \mathbf{ k \cdot v }\right) \mathbf{+}\omega _{ce}\frac{%
	\partial }{\partial \varphi _{v}}+\omega _{ge}\frac{\partial }{\partial
	\varphi _{s}}\right] \tilde{f}_{1}  
	= 
	\left[ \frac{q_e}{m_e}\tilde{\mathbf{E}}
	+\frac{i \mu _{e}}{m_e}\left(  \hat{\mathbf{s}} \cdot \tilde{\mathbf{B}}_{1} 
	+ \tilde{\mathbf{B}}_{1}\cdot \nabla_{\hat{ \mathbf{s} }}\right)  \mathbf{k} \right] 
	\cdot \nabla_\mathbf{v}f_{0} + \frac{2\mu _{e}}{\hbar }
	( \hat{\mathbf{s}} \times \tilde{\mathbf{B}}_{1}) \cdot\nabla_{\hat{ \mathbf{s}} } f_{0}   	
	\label{ansatz-lin-long}
\end{split}
\end{equation}
where we have introduced $\omega _{ge}=2\mu _{e} B_{0}/\hbar $ and $\omega _{ce}= q_e B_{0}/m_e$. We note that $\omega _{ge}=(g/2-1)\omega _{ce}$. Following Ref. \cite{newarticle}, the above equation can be solved
by an expansion in the eigenfunctions 
\begin{equation}
\psi _{n_1}(\varphi _{v},v_{\bot })=\exp [-i(n_1 \varphi _{v}-k_{\bot }v_{\bot
}\sin \varphi _{v}/\omega _{c})] ,
\end{equation}
where we use cylindrical coordinates for the velocity $\mathbf v = (v_\perp \cos \varphi_v, 
v_\perp \sin \varphi_v, v_z)$.
Thus, we let 
\begin{equation}
	\tilde{f}_{1}=\sum_{n_1,n_2}g_{n_1 n_2}(v_{\bot },v_{z},\theta _{s})\psi_{n_1}
	(\varphi _{v},v_{\bot })\exp (-i n_2\varphi _{s}) , 
	\label{expand}
\end{equation}
where $n_1=0,\pm 1,\pm 2,...$ and $n_2=-1,0,1$, where we have used spherical coordinates for the spin $\hat{\mathbf s} = (\cos \theta_s \sin \varphi_s, \sin \theta_s \sin \varphi_s, \cos \theta_s)$. The expansion above could
contain any integer $m$, but it so happens that after integration over $\varphi _{s}$, only $n_2=-1,0,1$ get a nonzero contribution,
as can be seen below. 
Using the orthogonality properties 
\[
	\frac{1}{2\pi} \int_{0}^{2\pi }\psi _{n}\psi _{m}^{\ast }d\varphi _{v}=\delta _{nm}
\]
we find 
\begin{equation}
	i\left( \omega -k_{z}v_{z}-n_1 \omega _{ce}-n_2 \omega _{ge}\right) 
	g_{n_1 n_2}=I_{n_1 n_2}(v_{\bot },v_{z},\theta _{s})  
	\label{resonance}
\end{equation}
with 
\begin{equation}
\begin{split}
	&
	I_{n_1 n_2}= 
	 \frac{1}{4\pi^2} 
	\int_{0}^{2\pi }\int_{0}^{2\pi }
	\left\{ \left[ \frac{q}{m} \tilde{\mathbf{E}} 
	+ \frac{i\mu _{e}}{m} \left( \hat{\mathbf{s}}  \cdot \tilde{\mathbf{B}}_{1}
	+ \tilde{\mathbf{B}}_{1}\cdot \nabla_{\hat{\mathbf s}} \right) \mathbf{k} \right] 
	\cdot \nabla_{\mathbf v} f_0 +\frac{2\mu _{e}}{\hbar }
	\left( \hat{\mathbf{s}} \times \tilde{\mathbf{B}}_{1}\right) \cdot 
	\nabla_{\hat{\mathbf s}} \right\} 
	\\ &\qquad
	\times \psi _{n_1}^{\ast }\exp (in_2\varphi _{s})\,d\varphi_{v}\,d\varphi _{s}  \label{coeff-1}
\end{split}
\end{equation}
A relation that is useful when trying to write results in a more explicit form is the Bessel-expansion 
\begin{equation}
	\psi _{n}(\varphi _{v},v_{\perp })=\sum_{m}J_{m}\left( \frac{k_{\bot }
	v_{\perp}}{\omega _{c}}\right) \exp [i(m-n)\varphi _{v}]  
	\label{Bessel-exp}
\end{equation}
Here it is seen that the results are much simplified in the limit where $k_{\perp}v_{te}/\omega _{c}$ is small (where we estimate $v_{\perp }$ with the thermal velocity $v_{te} = \sqrt{ k_B T_e / m_e} $), but in general the conductivity tensor turns into a sum over various combinations of Bessel functions. The conductivity tensor $\bar{\sigma}_{ij}$ for each species $s=e,i$, as defined by 
\begin{equation}
	j_{(s)}^{i}=\bar \sigma _{(s)}^{ij}E_{j} ,
\end{equation}
is found from (Eq.\ \eqref{currentdensity})
\begin{equation}
\mathbf{j}_{(s)} = q_{s}\int d^3v\,d^2\hat{s}\,\mathbf{v}f_{1s} + 3 \mu _{gs}\nabla \times \left( \int d^3v\,d^2\hat{s}\,\hat{ \mathbf{s} }%
	f_{1s}\right)   
	\label{magn-lin}
\end{equation}
by expressing the magnetic field in terms of $\tilde{\mathbf{E}}$, and then solving for $\tilde{f}_{1}$ in terms of $\tilde{\mathbf{E}}$ using the eigenfunctions as outlined above. For the ions the second term, i.e.\ the magnetization part is negligible due to the small magnetic moment of the ions. Similarly for the ion correspondence of Eq.\ \eqref{lin-long}, all spin terms are neglected and thus the classical Vlasov equation is used. So far the linear theory presented here applies to the general case. Without loss of generality we can let $\mathbf{k} = k_{\perp }\hat{\mathbf{x}}+k_{z}\hat{\mathbf{z}}$. We will now focus on a specific geometry.  For the specific case of shear Alfv\'{e}n waves we may have the approximate polarization $\tilde{\mathbf{E}}=\tilde{E}_{x}\hat{\mathbf{x}}$ and $\tilde{\mathbf{B}}_{1}=\tilde{B}_{y}\hat{\mathbf{y}}$, in
which case the (approximate) dispersion relation reads
\[
	\omega ^{2}-k_{z}^{2}c^{2} + \frac{i\omega }{\varepsilon _{0}}\sum_{s}\bar\sigma_{(s)}^{xx}=0 ,
\]
where Ampere's law has been used. 
That this polarization is indeed possible must be checked evaluating the
full linear theory involving all components of $\bar \sigma ^{ij}$ \cite{swanson}. In a regime without very high
temperatures or low temperatures, we may expect the standard classical
Vlasov theory \ to be applicable to a first approximation. However, as is
evident from (\ref{resonance}), the introduction of spin gives new
resonances, which may significantly affect the resonant wave-particle interaction even if the spin terms are otherwise small. For shear
Alfv\'{e}n waves $\omega \ll \omega _{ci}$, where $\omega _{ci}$ is the ion
cyclotron frequency. Making an expansion in $\omega /\omega _{ci},$ the
standard classical theory shows that it is the $g_{10}$-term for the ions
(recall that ions are always classical with $n_2=0,$ since their magnetic
moment is negligible) that gives the dominate contribution to the current, 
since the electron terms scales as  $\omega /\omega _{ce}$, compared to $%
\omega /\omega _{ci}$ for ions. However, the possibility to have spin-terms $%
g_{1-1}$ and $g_{-11}$ in the expansion opens up for the electron
contribution to be significant even in the regime $\omega \ll \omega _{ci}$,
since the factor $\left( \omega -k_{z}v_{z}-n_1\omega _{c}-n_2\omega
_{cg}\right) $ becomes reduced for the cases $(n_1, n_2)=(1, -1)$ and $(n_1, n_2) = (-1, 1)$.
In particular, wave particle corresponding to these terms may occur in the
bulk of the thermal distribution rather than in the exponentially small
tail. Thus when computing $\bar \sigma ^{xx}$ we keep the terms that are
dominating classically, which is $g_{10}$ and $g_{-10}$ (given that the
classical $g_{00}$-terms does not contribute to $\bar\sigma ^{xx}$, see e.g.
Ref.\ \cite{swanson}), together with the $g_{1-1}$ and $g_{-11}$ terms for electrons.
After straightforward algebra, assuming $\omega ^{2}\ll k_{z}^{2}c^{2}$ the
result is    
\begin{equation}
\begin{split}
	&
	k_{z}^{2}c^{2}
	+ \sum_{n_1=\pm 1,s=i,e} \omega _{ps}^{2}\int d^3v\,d^2\hat{s}\, \frac{\omega }{%
	(\omega -k_{z}v_{z} - n_1\omega _{cs})} 
	\left( \frac{\omega_{cs}}{k_\perp v_\perp} \right)^2 J_{n_1}^{2} 
	\left( \frac{k_{\bot }v_{\perp }}{\omega _{cs}}\right) \tilde{f}_{0} 
\\ & \qquad 
	 - \frac{3\omega_{pe}^2  k_z^2 \hbar^2 }{8 m_e^2} \sum_{n=\pm 1}  \int d^3v\,d^2\hat{s}\,
	\frac{ n \sin^2 \theta_s }{\omega - k_z v_z - n (\omega_c -\omega_{cg}) } 
	J^2_n \left( \frac{k_\perp v_\perp}{\omega_c }\right)
	\left[   \frac{\omega_c}{v_\perp} \frac{\partial \tilde{f}_0}{\partial v_\perp} 
	+ n k_z \frac{\partial \tilde{f}_0}{\partial v_z} 
	- \frac{2 m_e}{\hbar \sin \theta_s}  \frac{\partial \tilde{f}_0}{\partial \theta_s}  \right]    = 0   ,
\label{DR-first}
\end{split}
\end{equation}
where we have normalized the distribution functions so that  $f_0 = n_0\tilde{f}_0$ where $n_0$ is the unperturbed number density, and introduced the plasma frequency for each species $\omega_{ps} = n_0 q_s^2 / \epsilon_0 m_s$. A number of simplifications can be made. Firstly, in the sum over the species, only the ions need to be included. Secondly, for $k_{\bot
}v_{th}/\omega _{cs}\ll 1$ we may use Taylor-expansion of the Bessel-functions. 
Thirdly, for the quantum term only the two pole contributions are kept, and we assume for simplicity that the resonant electron velocity can be approximated as $v_{\textrm{res}} \equiv (\omega - \Delta \omega_{ce}) / k_z \approx \Delta \omega_{ce}/k_z$, where $\Delta \omega_{ce} \equiv \omega_{ce} - \omega_{cg}$. 
Provided that the wave frequency is approximately real, the dispersion relation then simplifies to 
\begin{equation}
k_{z}^{2}c^{2}-\omega _{pi}^{2}\left[ \frac{\omega ^{2}}{\omega _{ci}^{2}}+%
\frac{i\pi \omega }{k_{z}v_{ti}}\exp\left(-\frac{\omega _{ci}^{2}}{k_{z}^{2}v_{ti}^{2}}\right)%
\right] 
+\frac{3 i \pi}{4} \frac{k_{z}\hbar ^{2} \omega _{pe}^{2} }{m_{e}^{2}v_{te}}\frac{%
k_{\bot }^{2}\omega }{\omega _{ce}^{2}}\exp\left(- \frac{\Delta \omega
_{ce}^{2}}{k_{z}^{2}v_{te}^{2}}\right)  = 0 , \label{DR-second}
\end{equation}
where the first imaginary term is the classical ion contribution, and the
second imaginary term is the spin contribution from the electrons. Neglecting
the damping we thus have the standard shear Alfv\'{e}n wave dispersion
relation, $\omega ^{2}=k_{z}^{2}c_{A}^{2}$, with the Alfv\'{e}n velocity
given by $c_{A}=c\omega _{ci}/\omega _{pi}$. For parameter values
corresponding to typical classical plasmas, the coefficient of the second
exponential is much smaller than that of the first exponential term.
However, since the quantum term can have a much small exponent, since the
resonance may lie in the bulk of the distribution at the same time as the
classical resonance lie in the tail, the spin term can be the dominating
wave damping mechanism in parts of wave number space. An example for
specific plasma parameters is given in Fig.\ 1. Here we have introduced the growth rate $\gamma = \textrm{Im} (\omega) = \textrm{Im} (\omega_{\rm cl} + \omega_{\rm sp}) = \gamma_{\rm cl} + \gamma_{\rm sp}$, with the classical and quantum contributions $\gamma_{\rm cl}$ and $\gamma_{\rm sp}$ to the growth
rate of (\ref{DR-second}), respectively. 

\begin{figure}
	\begin{center}
	\includegraphics[width=.7\columnwidth]{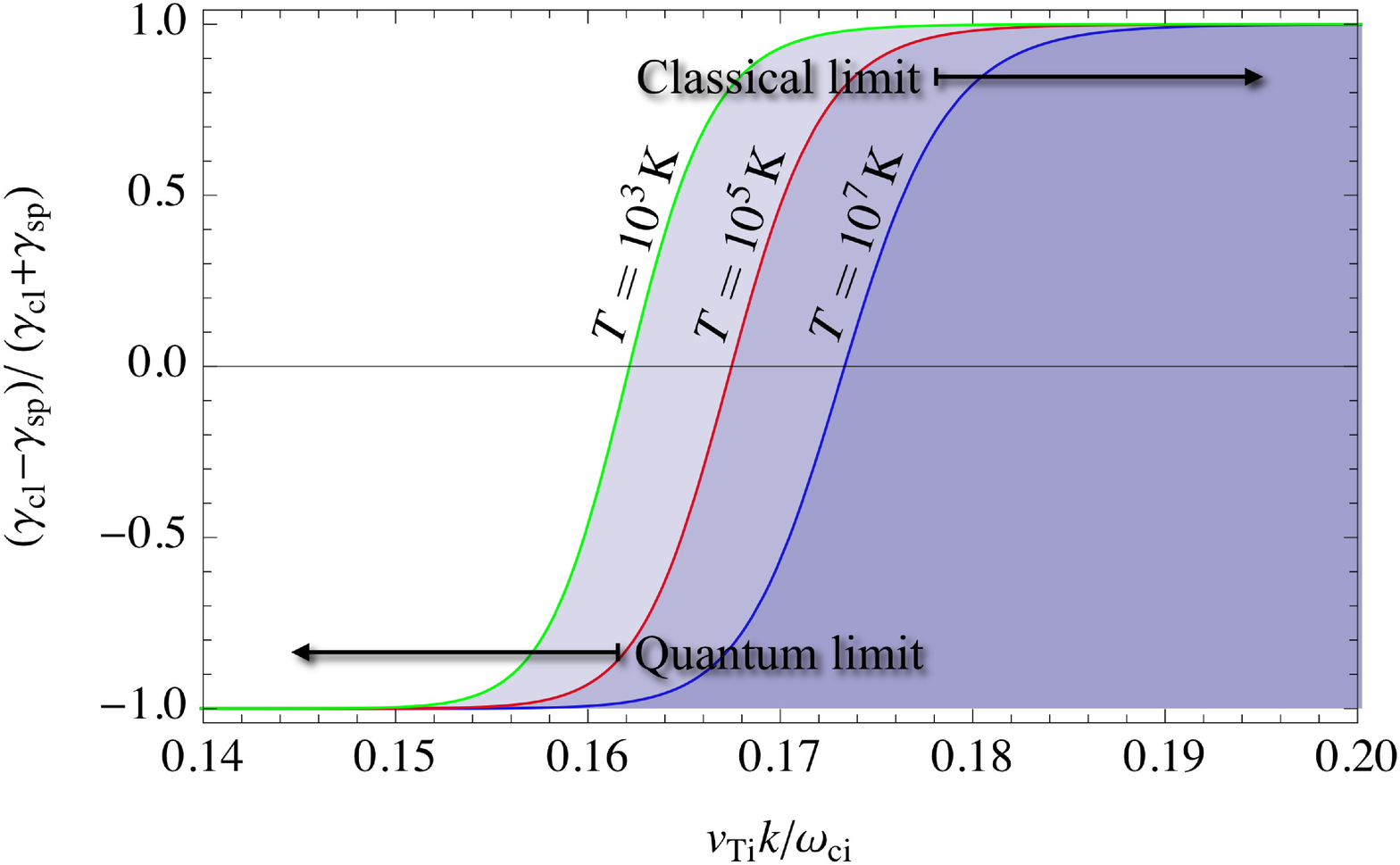}
	\end{center}
	\caption{
	The dependence of the normalized growth rate $\overline{\gamma }=\left( \gamma _{\mathrm{cl}}-\gamma _{\mathrm{sp}}\right) /(\gamma _{\mathrm{cl}}+\gamma _{\mathrm{sp}})$ on the normalized wavenumber $\overline{k}=k_{z}v_{thi}/\omega _{ci}$, for $n_{0}=10^{24}$ $\mathrm{m}^{-3}$, $B_{0}=10$ $\mathrm{T}$ and $k_{\bot }=3\times 10^{6}\mathrm{m}^{-1}$ in a plasma with equal electron and ion temperatures, $T_{e}=T_{i}=T$. It is clear that $\overline{\gamma}\rightarrow -1$ in the spin dominated damping regime to
the left and $\overline{\gamma }\rightarrow 1$ in the classically dominated regime to the right. Besides depending on the normalized wavenumber, the transition from
quantum to classical cyclotron damping depends slightly on the temperature, and the temperatures chosen here are $T=10^{3}\,\mathrm{K}$, $10^{5}\,\mathrm{K}$ and $10^{7}\,\mathrm{K}$. As a specific example we note that for $\overline{k}=0.15$ and $T=10^{5}$ $\mathrm{K}$, the damping is essentially due to the spin, i.e. $\gamma \approx \gamma _{\mathrm{sp}}$, with $\gamma _{\mathrm{sp}}=0.1\mathrm{rad/s}$.
}
\end{figure}

\subsection{Generalized $L$ and $R$ waves}

As a further example in the linear regime we linearize the full evolution equation, Eq.\ \eqref{eq:full-wigner-equation-2}. To simplify the algebra we look at waves propagating parallel to a background magnetic field. 
Following the steps from the last subsection it is straightforward to derive the dispersion relation
\begin{equation} \label{77}
	\det \left( \left[ 
	\begin{array}{ccc}
		 \omega^2 -  k^2 c^2  & 0 & 0 \\
		0 &\omega^2 -  k^2 c^2 & 0 \\ 
		0 & 0 & \omega^2 
	\end{array}
	\right]
	+ \frac{ i \omega}{\epsilon_0} \bar \sigma \right) = 0, 
\end{equation}
where
\begin{equation}
\begin{split}
&	
	\bar \sigma_{xx} = \bar \sigma_{yy} = \sum_{\pm} \int d\Omega 
	\frac{1}{\omega \mp \omega_c - k v_z} 
	\Bigg\{ - \frac{i q^2}{2m\omega} (k v_z - \omega)  
	\mp \frac{i q^3 B}{2m^2 \omega} \left[ 1 - \cos\left( \frac{i\hbar k}{2m} \partial_{v_z}  
	\right) \right] 
\\ & \quad
	- \frac{q^2 v_\perp^2}{2\hbar \omega} \sin \left( \frac{ i \hbar k}{2m} \partial_{v_z} \right)   
	+ \frac{3k^2 \mu_B^2}{\hbar \omega} \left[ \cos 2\theta_s 
	\sin \left( \frac{i\hbar k}{2m} \partial_{v_z} \right)
	\pm i \cos \theta_s \cos \left( \frac{i\hbar k}{2m} \partial_{v_z} \right) \right] 
	\Bigg\} f_0 ,
\\ &
	\bar\sigma_{xy} = - \bar \sigma_{yx} = \sum_{\pm} \int d\Omega 
	\frac{1}{\omega \mp \omega_c - k v_z} 
	\Bigg\{ \pm \frac{q^2}{2m\omega} (k v_z - \omega) 
	+ \frac{q^3 B}{2m^2 \omega} \left[ 1 - \cos\left( \frac{i\hbar k}{2m} \partial_{v_z}  
	\right) \right] 
\\ & \quad
	\mp \frac{i q^2 v_\perp^2}{2\hbar \omega} 
	\sin \left( \frac{ i \hbar k}{2m} \partial_{v_z} \right) 
	+ \frac{3k^2 \mu_B^2}{\hbar \omega} \left[ \pm i \cos 2\theta_s 
	\sin \left( \frac{i\hbar k}{2m} \partial_{v_z} \right)
	- \cos \theta_s \cos \left( \frac{i\hbar k}{2m} \partial_{v_z} \right) \right]   
	\Bigg\} f_0 ,
\\ &
	\bar\sigma_{zz} =  - \frac{ q^2}{\hbar k}\sum_{\pm} \int d\Omega \frac{1}{\omega - kv_z} 
	 \sin \left( \frac{i\hbar k}{2m} \partial_{v_z} \right) f_0 ,
\end{split}
\end{equation}
and $\bar\sigma_{xz} = \bar\sigma_{yz} = \bar\sigma_{zx} = \bar\sigma_{zy} = 0$. 
This dispersion relation reduces to the classical dispersion relation for \textit{L} and \textit{R} waves in the limit $\hbar \rightarrow 0$. This dispersion relation clearly shows the contribution from the spin as well as the the particle dispersive that becomes significant in the short wavelength regime.

A thorough discussion of the dispersion relation \eqref{77} is beyond the scope of the present paper. However, a few things can be noted on dimensional grounds. Firstly, we see that the higher order terms in the sin- and cos- operators become important for $k \Lambda_{dB} \sim 1$, where $\Lambda_{dB} \equiv \hbar / (m_e v_{te})$ is the thermal de Broglie wavelength for the electrons. For collective effects to be significant for such short wavelengths, we need $\hbar \omega_{pe}/(k_B T_e) \sim 1$. Secondly, quantum effects associated with the zero order distribution function tend to be significant if either $\mu_{e} B/(k_B T_e) \sim 1$ (Landau quantization and unsymmetric spin populations \cite{landau}) or if $\hbar^2 n_0^{2/3}/(m_e k_B T_e) \sim 1$ (Fermi-Dirac rather than Maxwell-Boltzmann statistics). Finally, the spin terms of Eq.\ \eqref{77} tend to be important in the regime $\hbar^2 \omega_{pe}^2/mc^2 k_B T_e \sim 1$. It should be stressed that these estimates may very well have to be revised when a thorough analysis is made, due to e.g.\ resonance effects.

\section{Gauge dependence} 

The definition of the Wigner function \eqref{4}  is not gauge invariant since it is a function of the gauge dependent canonical momentum rather than the gauge independent kinetic momentum $m\mathbf v = \mathbf p - q \mathbf A(\mathbf x)$. The theory above is hence only valid in the Coulomb gauge. It is possible to modify the definition to obtain a gauge independent Wigner function \cite{serimaa86}. 
In principle, there is nothing that prevents us to use a gauge dependent Wigner function as long as care is taken when doing gauge transformations. However, problems may arise when calculating for example the second order moment of the velocity $\left< \hat v_i \hat v_j \right>$. One might then be tempted to write
\begin{equation}
	\int d^3 x\, d^3 v\,  v_i v_j f(\mathbf x, \mathbf v, t) 
	= \int d^3 x\, d^3 p\, [ \hat p_i - q A_i (\hat{\mathbf x}) ] 
	[ \hat p_j - q A_j (\hat{\mathbf x}) ]. 
\end{equation}
However, the phase space function which is related to the operator $\hat v_i \hat v_j$ is \textit{not} $[ p_i - q  A_i (\mathbf x) ][p_j - q A_j (\mathbf x) ]$. In order to obtain the right function it is necessary to first put operator $[ \hat p_i - q A_i (\hat{\mathbf x}) ][ \hat p_j - q A_j (\hat{\mathbf x}) ]$ in Weyl-ordering \cite{weyl} and then make the substitution $\hat{\mathbf x} \rightarrow \mathbf x, \hat{\mathbf p} \rightarrow \mathbf p$. This is in general difficult to do since the vector potential is a function of $\mathbf x$. 
However, in the current paper we have only considered first order moments of the velocity and the ordering problem will not arise. We may hence use our distribution function to calculate for example the free charge current 
\begin{equation}
	\mathbf j_f ( \mathbf x, t) = \int d^3 v\, d^2 \hat s\, \mathbf v f(\mathbf x, \mathbf v, \hat{\mathbf s},t) 
	= \int d^3  p\ d^2 \hat s\ , \left[ \mathbf p - q \mathbf A(\mathbf x) \right]
	f(\mathbf x, \mathbf p, \hat{\mathbf s},t). 
\end{equation}
In agreement with Eq.\ \eqref{currentdensity} .

\subsection{Gauge invariant distribution function}

For completeness the fully gauge invariant distribution function is given here. Following Ref.\ \cite{serimaa86} the Wigner matrix is defined by
\begin{equation}
	W(\mathbf x, \mathbf v, \alpha, \beta, t) = 
	\frac{1}{(2\pi\hbar)^3} \int d^3 z \exp\left\{-\frac{im }{\hbar} \mathbf v \cdot \left[ \mathbf z 
	+q \int_{-1/2}^{1/2} d\tau \mathbf A(\mathbf x+ \tau \mathbf z, t) \right] \right\}
	\rho\left(\mathbf x + \frac{\mathbf z}{2} , \alpha;  
	\mathbf x - \frac{\mathbf z}{2}, \beta \right) . 
\end{equation}
where $\mathbf v$ is the velocity. The explicit dependence of the vector potential in this construction is there to compensate for the phase factor which the wave function acquires under a gauge transformation. Using the spin projection \eqref{9}, we obtain a fully gauge invariant distribution function. The evolution equation for the gauge invariant distribution function without spin was derived in Ref.\ \cite{serimaa86}. It is straightforward to generalize this equation to include spin with the result
\begin{eqnarray}
	\frac{\partial f}{\partial t} + (\mathbf v + \Delta \tilde{\mathbf v} ) \cdot \nabla_x f + 
	\frac{q}{m} \left[ (\mathbf v + \Delta \tilde{\mathbf v} ) \times \tilde{\mathbf B} + 
	\tilde{\mathbf E} \right] \cdot \nabla_v f + 
	\frac{\mu}{m} \nabla_x [ (\hat{\mathbf s} + \nabla_{\hat s} ) \cdot \tilde{\mathbf B} ] 
	\cdot \nabla_v f 
	+ \frac{2\mu}{\hbar} \left[ \hat{\mathbf s} \times
	\left(  \tilde{\mathbf B} + \Delta \tilde{\mathbf B} \right) \right] \cdot \nabla_{\hat s} f = 0, 
	\label{82}
\end{eqnarray}
where we have defined 
\begin{eqnarray}
	&&\!\!\!\!\!\!\!\!\!\!\!\!\!\!\!\!\!\!
	\tilde{\mathbf E} 
	= \int_{-1/2}^{1/2} d\tau \mathbf E \left(\mathbf x + \frac{i\hbar\tau}{m} 
	\nabla_v \right) 
	= \mathbf{E}(\mathbf{x}) \int_{-1/2}^{1/2}d\tau\, \cos\left(  \frac{\tau\hbar}{m}\stackrel{\leftarrow}{\nabla}_x\cdot
	 \stackrel{\rightarrow}{\nabla}_v\right)  
\\ &&\!\!\!\!\!\!\!\!\!\!\!\!\!\!\!\!\!\!
	\tilde{\mathbf B}  
	= \int_{-1/2}^{1/2} d\tau \mathbf B \left(\mathbf x + \frac{i\hbar\tau}{m} 
	\nabla_v \right) 
	= \mathbf{B}(\mathbf{x})\int_{-1/2}^{1/2}d\tau\, \cos\left(  \frac{\tau\hbar}{m}\stackrel{\leftarrow}{\nabla}_x\cdot
	 \stackrel{\rightarrow}{\nabla}_v\right)  
\\ &&\!\!\!\!\!\!\!\!\!\!\!\!\!\!\!\!\!\!
	 	\Delta \tilde{\mathbf v} 
		= - \frac{iq\hbar}{m^2} \int_{-1/2}^{1/2} 
	d\tau\, \tau \mathbf B \left( \mathbf x + \frac{i \hbar \tau}{m} \nabla_v  \right) \times \nabla_v 
	= 
	\frac{q\hbar}{m^2}\left[ \mathbf B(\mathbf x) \int_{-1/2}^{1/2} 
	d\tau\, \tau \sin\left( \frac{\tau\hbar}{m}\stackrel{\leftarrow}{\nabla}_x\cdot
	 \stackrel{\rightarrow}{\nabla}_v \right)\right] \times \stackrel{\rightarrow}{\nabla}_v 
\\ &&\!\!\!\!\!\!\!\!\!\!\!\!\!\!\!\!\!\!
	\Delta \tilde{\mathbf B} 
	= - \frac{i\hbar}{m} \int_{-1/2}^{1/2} d\tau\, \tau 
	\mathbf B \left( \mathbf x + \frac{i\hbar\tau}{m}\nabla_v \right)  
	 \stackrel{\leftarrow}{\nabla}_x\cdot \stackrel{\rightarrow}{\nabla}_v 
	 =
	 \frac{\hbar}{m}\mathbf{B}(\mathbf{x})\int_{-1/2}^{1/2}d\tau\, \tau\sin\left( 
	 	\frac{\tau\hbar}{m}\stackrel{\leftarrow}{\nabla}_x\cdot \stackrel{\rightarrow}{\nabla}_v
	 \right)\stackrel{\leftarrow}{\nabla}_x\cdot \stackrel{\rightarrow}{\nabla}_v . 
\end{eqnarray}	
Thus, to lowest order in $\hbar$ we have 
\begin{eqnarray}
&&
	\tilde{\mathbf E} 
	\approx \mathbf{E}(\mathbf{x})\left[ 1 - \frac{\hbar^2}{24m^2}\left(\stackrel{\leftarrow}{\nabla}_x\cdot\stackrel{\rightarrow}{\nabla}_v\right)^2 \right]
\\ &&
	\tilde{\mathbf B}  
	\approx \mathbf{B}(\mathbf{x})\left[ 1 - \frac{\hbar^2}{24m^2}\left(\stackrel{\leftarrow}{\nabla}_x\cdot\stackrel{\rightarrow}{\nabla}_v\right)^2 \right]
\\ &&
	\Delta \tilde{\mathbf v} 
	\approx  \frac{q\hbar^2}{12m^3}\mathbf B(\mathbf{x}) \times \stackrel{\rightarrow}{\nabla}_v \left(\stackrel{\leftarrow}{\nabla}_x\cdot\stackrel{\rightarrow}{\nabla}_v\right) ,  
\\ &&
	\Delta \tilde{\mathbf B} 
	\approx \frac{\hbar^2}{12m^2}  
	\mathbf B(\mathbf{x})  
	 \left(\stackrel{\leftarrow}{\nabla}_x\cdot \stackrel{\rightarrow}{\nabla}_v \right)^2
	,
\end{eqnarray}	
so that (cf. (40))
\begin{equation}
\begin{split}
&
	\frac{\partial f}{\partial t} + \mathbf v \cdot \nabla_x f + 
	\left[ 
		\frac{q}{m}(\mathbf{E} + \mathbf v\times \mathbf{B})\cdot \nabla_v + 
		\frac{\mu_B}{m} \nabla_x [ (\hat{\mathbf s} + \nabla_{\hat s} ) \cdot \mathbf{B} ] 		\cdot \nabla_v + \frac{2\mu_B}{\hbar}(\hat{\mathbf{s}}\times\mathbf{B})\cdot\nabla_{\hat{s}}
	\right] f
\\ &	 
	 = \frac{\hbar^2}{24m^2}\Bigg\{ 
	 		\left[ 
		\frac{q}{m}(\mathbf{E} + \mathbf v\times \mathbf{B})\cdot \nabla_v + 
		\frac{\mu_B}{m} \nabla_x [ (\hat{\mathbf s} + \nabla_{\hat s} ) \cdot \mathbf{B} ] 		\cdot \nabla_v - \frac{2\mu_B}{\hbar}(\hat{\mathbf{s}}\times\mathbf{B})\cdot\nabla_{\hat{s}}
	\right]\left(\stackrel{\leftarrow}{\nabla}_x\cdot\stackrel{\rightarrow}{\nabla}_v\right)^2 
\\ & \qquad
	- 2\left[ \frac{q}{m}
	 	\mathbf B\times \nabla_v \left(\stackrel{\leftarrow}{\nabla}_x\cdot\stackrel{\rightarrow}{\nabla}_v\right )
		\right] \cdot
		\left( 
		\frac{q}{m}\mathbf{B}\times\nabla_v + \nabla_x \right)
		\Bigg\}f.
\end{split}
\end{equation}
The gauge invariant Wigner function has a modified Weyl correspondence which is well suited for calculating fluid moments. In order to obtain the phase space $O(\mathbf x, \mathbf v)$ function which corresponds to an operator $O(\hat{\mathbf x}, \hat{\mathbf v})$, all products of the operators $\hat{\mathbf x}$ and $\hat{\mathbf v} \equiv [\hat{\mathbf p} - q \mathbf A(\hat{\mathbf x})]/m$ are first ordered in a symmetric form using the commutation relation $\hat{\mathbf x}$ and $\hat{\mathbf p}$ and then the substitution $\hat{\mathbf x} \rightarrow \mathbf x$ and $\hat{\mathbf v} \rightarrow \mathbf v$ is taken (details can be found in \cite{serimaa86}). 

\begin{figure}
	\begin{center}
	\includegraphics[width=.7\columnwidth]{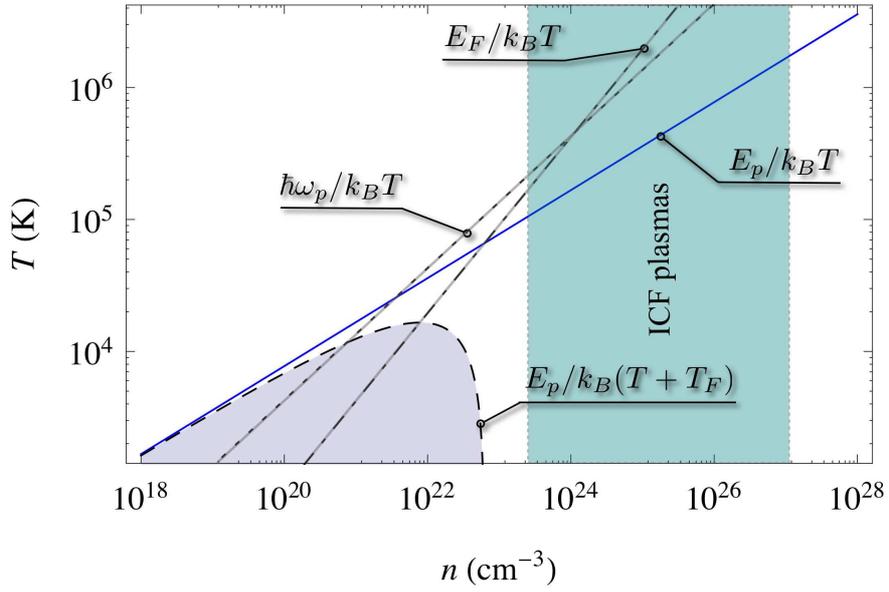}
	\end{center}
	\caption{Various plasma regimes in the temperature-density parameter space
are illustrated. The dotted line is given by the strong coupling
parameter $\Gamma =E_{p}/k_{B}T=1$, where $E_{p}=e^{2}n_{0}^{1/3}/4\pi \varepsilon _{0}$
is the potential energy due to the nearest neighbor. For larger densities
this parameter is repaced by $\Gamma _{F}=E_{p}/k_{B}(T+T_{F})$, since the
average kinetic energy of the particles is given by the Fermi energy rather
than the thermal energy. The curve $\Gamma _{F}=1$ is illustrated by the
dashed curve, and the strong coupling region, which is shaded, occurs below
this line. In this region our model is not directly applicable, since collisions has
not been taken into account. For comparison we have also drawn the lines
$\hbar \omega _{p}/k_{B}T$ (the dotted grey line) and the line $T_{F}/T$ (the dotted-dashed grey line)
which measures the importance of wave function dispersion and the Fermi
pressure, respectively. As a rough estimate, the quantum regime is below
either of these lines. Note, however, that spin effects can sometimes be important even
above these lines \cite{Brodin-Marklund-Manfredi}}
\end{figure}

\section{Summary and discussion}\label{sec:sum}

In the present paper we have derived an evolution equation, Eq. \eqref{eq:full-wigner-equation-2}, for a
quasi distribution function of electrons, based on a Wigner transformation
of the density matrix, together with a spin operator contracting the $2\times 2$ 
Wigner-matrix to a scalar function  $f(\mathbf{x}, \mathbf{p}, \mathbf{s})$. The free current and the magnetization can be directly computed from the quasi distribution
function, and hence Eq.\ \eqref{eq:full-wigner-equation} (or the gauge invariant alternative, Eq.\ \eqref{82}) together with Maxwell's equations with the sources \eqref{chargedensity}, and \eqref{currentdensity}, form a closed set.
The present theory has the advantage to include the full
quantum dynamics in a single equation, and provide an immediate path
between the classical and quantum descriptions. For macroscopic scale
lengths longer than the characteristic de Broglie wavelength, the kinetic
equation greatly simplifies. In particular, the semi-classical kinetic
theory put forward in Ref.\ \cite{newarticle} is recovered, but with some small but
significant deviations as shown in \eqref{eq:semi-classical}. 
The difference between the semi-classical theory and our result follows from the smeared out probability distribution of the spin.

In order to illustrate the theory, examples of linear wave propagation
solving the full quantum theory, Eq. \eqref{eq:full-wigner-equation-2}, as well as the long wavelength 
limit, Eq. \eqref{eq:semi-classical} are given. An interesting result is that the wave damping
can be much affected by the non-classical terms even in a supposedly
classical temperature and density regime, although the real part of the
wave frequency then is always well approximated by the classical Vlasov
theory. The reason is that the spin terms give raise to new types of wave
particle resonances.

Proper initial conditions for the quasi distribution function can be found
by computing the Wigner transformation of the density matrix in the
thermodynamical ground state. The result for the simple but important
special case of a magnetic field is given, see Eq. \eqref{eq:equilibrium}, in which case the
energy levels are Landau quantized and split due to the two spin states \cite{landau}.

The present quantum theory can be used for a broad range of parameters,
extending the applicability to regimes of high densities, strong magnetic
fields, low temperatures and short scalelengths, that are not covered by
the classical Vlasov equation. However, there is still much room for
improvements. In particular, when the strong coupling parameter $\Gamma$ is
increased, collisional effects become important \cite{fortov}. A schematic view of the different plasma regimes is given in Fig.\ 2. Furthermore, the present theory does not account for relativistic effects that is crucial in e.g.\ laser plasma interaction. Removal of these restrictions, as well as a more complete evaluation of the present theory, constitutes interesting projects for future research.

\section*{Acknowledgments}
This research is supported by the European Research Council under Constract No.\ 204059-QPQV and  the Swedish Research Council under Contracts No.\ 2007-4422.

\end{document}